\documentclass[fleqn,usenatbib]{mnras}

\usepackage{newtxtext,newtxmath}

\usepackage[T1]{fontenc}

\DeclareRobustCommand{\VAN}[3]{#2}
\let\VANthebibliography\thebibliography
\def\thebibliography{\DeclareRobustCommand{\VAN}[3]{##3}\VANthebibliography}

\usepackage{graphicx}	
\usepackage{xcolor}
\usepackage[normalem]{ulem}
\usepackage{amsmath}	
\usepackage{orcidlink}
\usepackage{pdflscape}  
\usepackage{threeparttable}  
\usepackage{booktabs}
\usepackage{caption}
\usepackage{orcidlink}

\newcommand{\HeII}{He\,\textsc{ii}\,$\lambda$4687}
\newcommand{\OII}{[O\,\textsc{ii}]}
\newcommand{\OIIdoublet}{[O\,\textsc{ii}]\,$\lambda\lambda$3727, 3730}
\newcommand{\Hb}{H$\beta$}
\newcommand{\Ha}{H$\alpha$}
\newcommand{\Hg}{H$\gamma$}
\newcommand{\NII}{[N\,\textsc{ii}]}
\newcommand{\NIIdoublet}{[N\,\textsc{ii}]\,$\lambda\lambda$6550, 6585}
\newcommand{\NIIleft}{[N\,\textsc{ii}]\,$\lambda$6550}
\newcommand{\SII}{[S\,\textsc{ii}]}
\newcommand{\SIIdoublet}{[S\,\textsc{ii}]\,$\lambda\lambda$6718, 6733}
\newcommand{\OIII}{[O\,\textsc{iii}]}
\newcommand{\OIIIdoublet}{[O\,\textsc{iii}]\,$\lambda\lambda$4960, 5008}
\newcommand{\OIIIARR}{[O\,\textsc{iii}]\,$\lambda$4364}
\newcommand{\NeIII}{[Ne\,\textsc{iii}]\,$\lambda$3870}

\newcommand{\CIVlambda}{C\,\textsc{iv}\,$\lambda$1549}
\newcommand{\HeIIlambda}{He\,\textsc{ii}\,$\lambda$1640}
\newcommand{\CIIIlambda}{C\,\textsc{iii}]\,$\lambda$1909}
\newcommand{\OIIIUVlambda}{O\,\textsc{iii}]\,$\lambda$1663}

\newcommand{\CII}{[C\,\textsc{ii}]}

\newcommand{\ergs}{\text{erg s$^{-1}$}}
\newcommand{\ergscm}{\text{erg s$^{-1}$ cm$^{-2}$}}

\newcommand{\kms}{\text{km s$^{-1}$}}

\graphicspath{{./}{figures/}}

\title[ALPINE-CRISTAL-JWST AGN]{The ALPINE-CRISTAL-JWST Survey: Revealing Less Massive Black Holes in High-Redshift Galaxies}

\author[Ren et al.]{
\parbox{\textwidth}{
\Large
Wenke Ren$^{1,2}$\thanks{E-mail: rwk@mail.ustc.edu.cn}\orcidlink{0000-0002-3742-6609},
John D. Silverman$^{1,3,4,5}$\orcidlink{0000-0002-0000-6977},
Andreas L. Faisst$^{6}$\orcidlink{0000-0002-9382-9832},
Seiji Fujimoto\thanks{Hubble Fellow}$^{7}$\orcidlink{0000-0001-7201-5066},
Lin Yan$^{8,9}$\orcidlink{0000-0003-1710-9339},
Zhaoxuan Liu$^{1,3,4,10}$\orcidlink{0000-0002-9252-114X},
Akiyoshi Tsujita$^{11}$\orcidlink{0000-0002-0498-5041},
Manuel Aravena$^{12,13}$\orcidlink{0000-0002-6290-3198},
Rebecca L. Davies$^{14}$\orcidlink{0000-0002-3324-4824},
Ilse De Looze$^{15}$\orcidlink{0000-0001-9419-6355},
Miroslava Dessauges-Zavadsky$^{16}$\orcidlink{0000-0003-0348-2917},
Rodrigo Herrera-Camus$^{17,13}$\orcidlink{0000-0002-2775-0595},
Edo Ibar$^{18,13}$\orcidlink{0009-0008-9801-2224},
Gareth C. Jones$^{19,20}$\orcidlink{0000-0002-0267-9024},
Jeyhan S. Kartaltepe$^{21}$\orcidlink{0000-0001-9187-3605},
Anton M. Koekemoer$^{22}$\orcidlink{0000-0002-6610-2048},
Yu-Heng Lin$^{6}$\orcidlink{0000-0001-8792-3091},
Ikki Mitsuhashi$^{23}$\orcidlink{0000-0001-7300-9450},
Juan Molina$^{18,13}$\orcidlink{0000-0002-8136-8127},
Ambra Nanni$^{24,25}$\orcidlink{0000-0001-6652-1069},
Monica Relano$^{26,27}$\orcidlink{0000-0003-1682-1148},
Michael Romano$^{28,29}$\orcidlink{0000-0002-9948-3916},
David B. Sanders$^{30}$\orcidlink{0000-0002-1233-9998},
Manuel Solimano$^{12}$\orcidlink{0000-0001-6629-0379},
Enrico Veraldi$^{31}$\orcidlink{0009-0007-1304-7771},
Vicente Villanueva$^{17}$\orcidlink{0000-0002-5877-379X},
Wuji Wang$^{6}$\orcidlink{0000-0002-7964-6749},
Giovanni Zamorani$^{32}$\orcidlink{0000-0002-2318-301X}
\\
{\it \footnotesize (Affiliations listed at the end of the paper)}
}
}

\date{Accepted XXX. Received YYY; in original form ZZZ}

\pubyear{\the\year{}}

\begin{document}
\label{firstpage}
\pagerange{\pageref{firstpage}--\pageref{lastpage}}
\maketitle

\begin{abstract}
	We present a systematic search for broad-line active galactic nuclei (AGNs) in the ALPINE-CRISTAL-JWST sample of 18 star-forming galaxies ($M_\star>10^{9.5}~M_{\odot}$) at redshifts $z=4.4-5.7$. Using JWST/NIRSpec IFU, we identify 7 AGN candidates through the detection of broad \Ha\ emission lines from 33 aperture spectra centred on photometric peaks. These candidates include one highly robust AGN detection with FWHM $\sim$ 2800 \kms\ and six showing broad components with FWHM $\sim 600-1600$ \kms, with two in a merger system. We highlight that only broad-line detection is effective since these candidates uniformly lie within narrow emission-line ratio diagnostic diagrams where star-forming galaxies and AGNs overlap. The broad-line AGN fraction ranges from 5.9\% to 33\%, depending on the robustness of the candidates. Assuming that the majority are AGNs, the relatively high AGN fraction is likely due to targeting high-mass galaxies, where simulations demonstrate that broad-line detection is more feasible. Their black hole masses range from $10^6$ to $10^{7.5}~M_{\odot}$ with $0.1 \lesssim L_{\rm bol}/L_{\rm Edd}\lesssim 1$. Counter to previous JWST studies at high redshift that found overmassive black holes relative to their host galaxies, our candidates lie close to or below the local $M_{\rm BH}-M_\star$ scaling relations, thus demonstrating the effect of selection biases. This study provides new insights into AGN-host galaxy co-evolution at high redshift by identifying faint broad-line AGNs in galaxy samples, highlighting the importance of considering mass-dependent selection biases and the likelihood of a large population of AGNs being undermassive and just now being tapped by JWST.      

\end{abstract}




\section{Introduction} \label{sec:intro}

Studies of local galaxies and Active Galactic Nuclei (AGNs) have revealed tight correlations between the masses of supermassive black holes (SMBHs; $M_{\rm BH}$) and their host galaxy properties, including velocity dispersion and stellar mass ($M_\star$) \citep{Ferrarese2000,Gebhardt2000,Haering2004,Kormendy2013,Reines2015,Greene2020,Ding2020}. These scaling relations suggest underlying physical mechanisms regulating galaxy-SMBH coevolution throughout cosmic history. Models and simulations, envisaging different possible coevolutionary (as well as non-coevolutionary) scenarios, end up with similar scaling relations as observed in the local universe \citep{Jahnke2011,King2015,Valiante2016,Weinberger2018,Habouzit2021,Koudmani2022}. A key to disentangle the coevolutionary scenario is to explore the population of AGNs at high redshift, which can provide strong constraints on the early assembly of SMBHs \citep[e.g.,][]{Inayoshi2020,Habouzit2021,Schneider2023}.

The James Webb Space Telescope (JWST) has dramatically enhanced our ability to study high-redshift AGNs. Over a hundred AGN candidates have been identified through JWST observations using multiple techniques, including broad-band spectral energy distribution (SED) fitting, narrow emission-line diagnostics, and X-ray observations \citep{Goulding2023,Kocevski2023,Kokorev2023,Onoue2023,Fujimoto2024,Mazzolari2024,Treiber2024,Uebler2024,Napolitano2025,Scholtz2025}. These groundbreaking discoveries are pushing the frontiers of AGN demographics to fainter luminosities and enabling detailed studies of the host galaxies of these early AGNs. A noteworthy finding from studies focusing on AGN samples with broad emission lines, which allow for the estimation of SMBH mass, is an apparent general trend of overmassive SMBHs at high redshift compared to the local $M_{\rm BH}-M_\star$ relation \citep{Harikane2023,Uebler2023a,Maiolino2024,Stone2024a,Yue2024b}.

These growing samples of high-redshift AGNs are subject to various selection biases. At higher black hole masses ($M_{\rm BH}>10^8~M_{\odot}$), studies targeting luminous quasars \citep{Ding2023,Stone2024a,Yue2024b} inherently select AGNs with overmassive black holes relative to their host galaxies \citep{Lauer2007}. Efforts to detect fainter AGNs through galaxy spectroscopic surveys \citep{Harikane2023,Maiolino2024}, which could probe AGNs at low black hole masses ($M_{\rm BH}\lesssim10^7~M_{\odot}$), face different challenges. As black hole mass decreases, broad emission lines become both fainter and narrower, making it increasingly difficult to distinguish AGN broad line regions (BLR) from outflow components and narrow emission lines. 

As a result, studies demonstrate that current samples are insufficient to place strong constraints on the intrinsic $M_{\rm BH}-M_\star$ relation at $z\gtrsim4$ with JWST, mainly limited by selection biases and measurement uncertainties \citep{Li2022,Li2024,Silverman2025}. To improve the current situation, \citet{Li2025} used a galaxy-based selection method to search for faint broad-line AGN signatures within a sample of massive galaxies, an approach that lessens the biases inherent in traditional AGN-luminosity-selected samples. This method revealed a population of 'normal-mass' black holes ($z\sim3-5$) -- those with mass ratios ($M_{\rm BH}/M_\star$) consistent with local scaling relations -- thus suggesting that selection biases in previous studies may have obscured the true diversity of the high-redshift black hole population. \citet{Geris2025} also detect faint broad \Ha\ lines by stacking $\sim$600 JWST JADES spectra, indicating a significant population of low-mass black holes ($\sim 10^6 M_\odot$) at high redshift.

To mitigate selection biases inherent in traditional AGN surveys, our study focuses on a well-defined sample of high-mass ($M_\star>10^{9.5}~M_{\odot}$) star-forming galaxies at $4<z<6$ from the ALPINE-CRISTAL-JWST program \citep[][submitted]{Faisst2025}. The parent sample was selected to be dominated by star formation, excluding known type-1 AGNs and luminous X-ray sources. While this pre-selection removes the most conspicuous active nuclei, the deep, spatially-resolved NIRSpec IFU observations from our program are crucial for detecting more subtle activity. These data enable us to effectively isolate potential nuclear emission from galactic outflows and host-galaxy contamination, facilitating a careful search for faint AGN signatures that might otherwise be overlooked. This approach complements existing high-redshift AGN samples and provides a more complete view of the broader AGN population.

This paper is organized as follows. Sec. \ref{sec:DandM} describes the data reduction and spectral fitting methodology. Sec. \ref{sec:detectability} explores the detectability of AGN signatures through simulations incorporating local scaling relations. In Sec. \ref{sec:BLAGN}, we present our AGN candidates and explain our selection criteria. Sec. \ref{sec:results} presents the measurements and properties of these candidates, including their positions on scaling relations and emission-line diagnostics. We discuss the implications of our findings in Sec. \ref{sec:discussion}, with particular focus on AGN fraction and gas kinematics. Sec. \ref{sec:summary} summarizes our conclusions. Throughout this paper, we assume a flat $\Lambda$CDM cosmology with $\Omega_{\rm m}=0.3$, $\Omega_{\Lambda}=0.7$, and $H_0=70~\kms~\text{Mpc}^{-1}$.

\begin{figure*}
   \includegraphics[width=\textwidth]{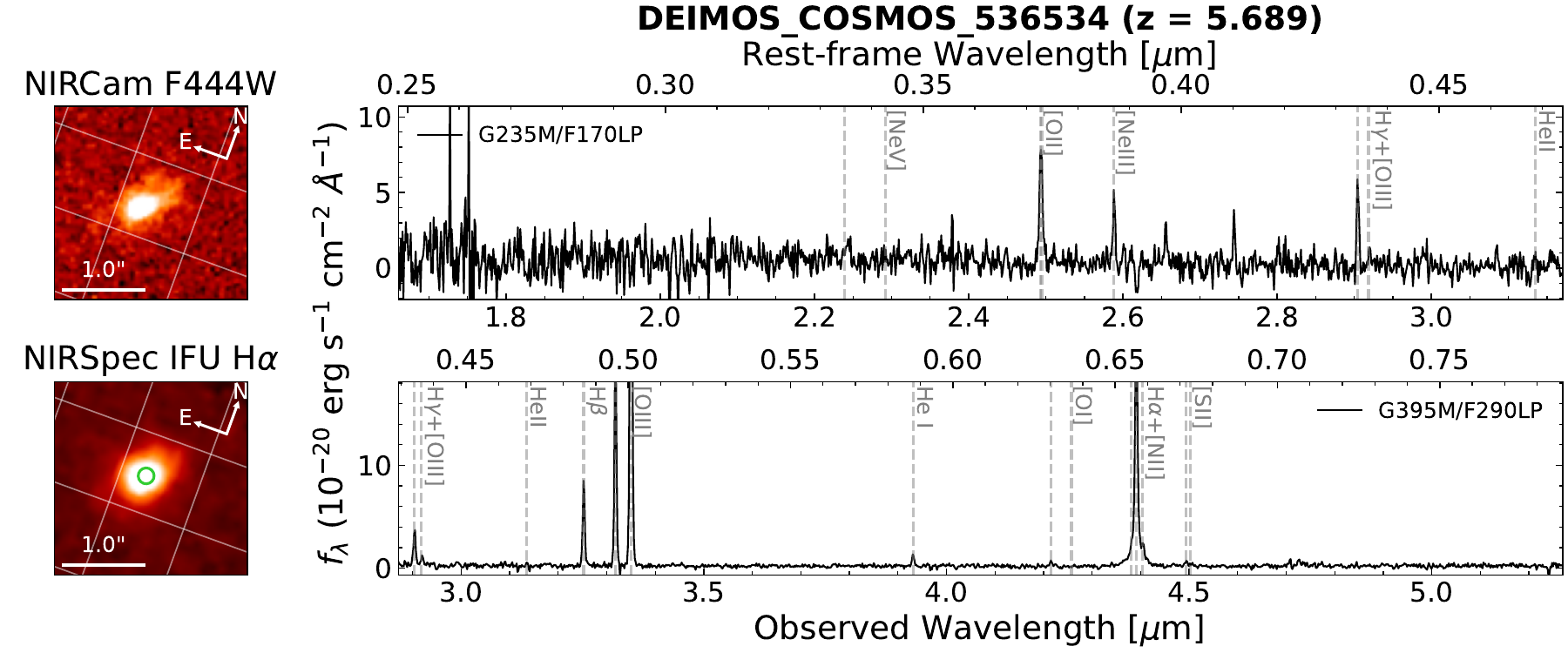}
   \caption{Demonstration of our aperture spectral extraction for \textit{DC\_536534}. Left: NIRCam F444W image (top) and the \Ha\ emission line flux map (bottom). For better visualization, we normalize the F444W image using an Asinh stretch and reproject the \Ha\ map to the world coordinate system (WCS) of the F444W image. The spectral extraction aperture is indicated by the green circle. Right: The extracted NIRSpec IFU spectrum from the marked aperture, showing key emission line features in both the G235M and G395M gratings.}
   \label{fig:aper_spec}
\end{figure*}

\section{Data and Methodology} \label{sec:DandM}

\subsection{Observations and Data Reduction} \label{subsec:obs}

The ALPINE-CRISTAL-JWST programme comprises ALMA and JWST observations of 18 representative main-sequence galaxies at $4.4<z<5.7$ \citep[JWST Cycle 2, GO PID: 3045,][submitted]{Faisst2025}. Selected to probe the star-forming main sequence at $z\sim 5$, these galaxies have stellar mass $M_\star>10^{9.5}~M_{\odot}$ and star formation rates SFR$>10~M_{\odot}$ yr$^{-1}$ ($M_{UV}<-20.2$). The stellar masses and SFRs are derived from SED fitting to archived photometry \citep{Faisst2020}. The final sample benefits from an extensive multi-wavelength dataset comprising high-resolution ALMA sub-millimetre observations, JWST NIRCam imaging, and NIRSpec IFU spectroscopy utilizing both G235M and G395M gratings (with \textit{DC\_842313} being the sole exception, lacking G395M coverage).

The ALMA observations are provided by two large surveys: ALPINE \citep[Cycle 5, Project ID: 2017.1.00428.L,][]{LeFevre2020} and CRISTAL \citep[Cycle 8, Project ID: 2021.1.00280.L,][]{HerreraCamus2025a}. The ALPINE survey examined 118 star-forming galaxies at $4 < z < 6$, located all in the COsmic Evolution Survey \citep[COSMOS,][]{Scoville2007} and in the Extended Chandra Deep Field South \citep[ECDFS,][]{Giacconi2002}. The ALMA project focuses on the \CII\ emission line to study the properties of the interstellar medium, including star formation, gas/dust content and kinematics \citep{Bethermin2020,Faisst2020}. The sample selection explicitly excluded known AGNs identified through X-ray observations or those exhibiting broad Ly$\alpha$ emission in their rest-frame UV spectra. The CRISTAL survey further targeted 19 galaxies from the ALPINE sample, aiming to achieve kiloparsec-scale resolution for detailed kinematic and morphological mapping of the gas component.

The JWST imaging data are from COSMOS-Web \citep{Casey2023}, a large programme in Cycle 1, designed to survey a 0.54 ${\rm deg}^2$ region of the COSMOS field with NIRCam imaging in F115W, F150W, F277W and F444W filters. In parallel, a smaller, non-contiguous 0.19 ${\rm deg}^2$ area is covered with MIRI imaging in the F770W filter.

Regarding the spectroscopic observations, we employed the NIRSpec IFU with both G235M and G395M gratings, providing continuous wavelength coverage from 1.66 to 5.10 $\mu$m at a spectral resolution of $R\sim1000$. This configuration enables detection of key rest-frame emission lines including \SIIdoublet, \NIIdoublet, \Ha, \OIIIdoublet, \Hb, and \OIIdoublet\ across our sample. We implemented a two-point dithering pattern to optimise sky subtraction and field of view coverage. The exposure times were determined using the Pandeia JWST Exposure Time Calculator (ETC). We estimated H$\alpha$ fluxes from the total SFRs of galaxies and subsequently derived the fluxes for the fainter [OII] and [NII] lines line ratios assuming conservative gas phase metallicities \citep{Maiolino2008} and dust obscuration \citep{Faisst2020}. These predicted fluxes, along with assumptions for source size, line width, and continuum magnitude, were input into the ETC to determine the exposure time required to achieve a predicted integrated S/N$>$5 for these key diagnostic lines.

We processed the data using the JWST Science Calibration Pipeline (v1.16.0), which includes detector-level corrections (bias subtraction, dark current removal, linearity correction), background subtraction using our 2-point dither pattern, wavelength and flux calibration, and data cube reconstruction. We enhanced the standard pipeline with improved sky subtraction algorithms and refined astrometric alignment procedures, as detailed in \citet[][in prep]{Seiji2025}. We adopt the nominal pixel scale of 0.1\arcsec\ for the NIRSpec IFU data and align its astrometry with the NIRCam imaging, whose absolute astrometry has been directly registered to a Gaia-aligned version of the original COSMOS HST ACS F814W imaging mosaics \citep{Koekemoer2007}. For the G235M data, we generate a synthetic F277W image and determine the astrometric offset by fitting a 2D elliptical Gaussian to both the IFU and NIRCam F277W images, comparing their centres and major axes. We apply an analogous procedure for the G395M data, generating a synthetic F444W image for comparison with the NIRCam F444W image. The final astrometric correction is derived from the average offset between these two measurements and applied uniformly to both G235M and G395M datasets.

To optimise detection of AGN features whilst minimizing host galaxy contamination, we extract aperture spectra for each target. Using the redshift determined from the far-infrared \CII\ line, we generate \Ha\ line flux maps (or \OIII\ for \textit{DC\_842313}) by collapsing the IFU data cube around rest-frame 6565 \AA\ (or 5008 \AA\ for \textit{DC\_842313}). We locate the optical centre of each flux map using the {\tt DAOFIND} algorithm \citep{Stetson1987} and extract spectra using a 0.2\arcsec-diameter aperture. This aperture size is selected to be marginally larger than the typical point spread function (PSF) FWHM of 0.16\arcsec\ in the F444W band where the \Ha\ line is situated \citep[e.g.,][]{DEugenio2024a}. Several of our targets are clumpy and/or merger systems \citep[e.g.,][]{LeFevre2020,Jones2021,Romano2021}, resulting in multiple centres in their \Ha\ line flux maps. Consequently, we identified a total of 33 apertures across our sample of 18 galaxies for AGN searches. Fig. \ref{fig:aper_spec} presents an example of these aperture spectra.

\subsection{Spectral Fitting: broad-line identification} \label{subsec:fitting}

\begin{figure}
   \includegraphics[width=\columnwidth]{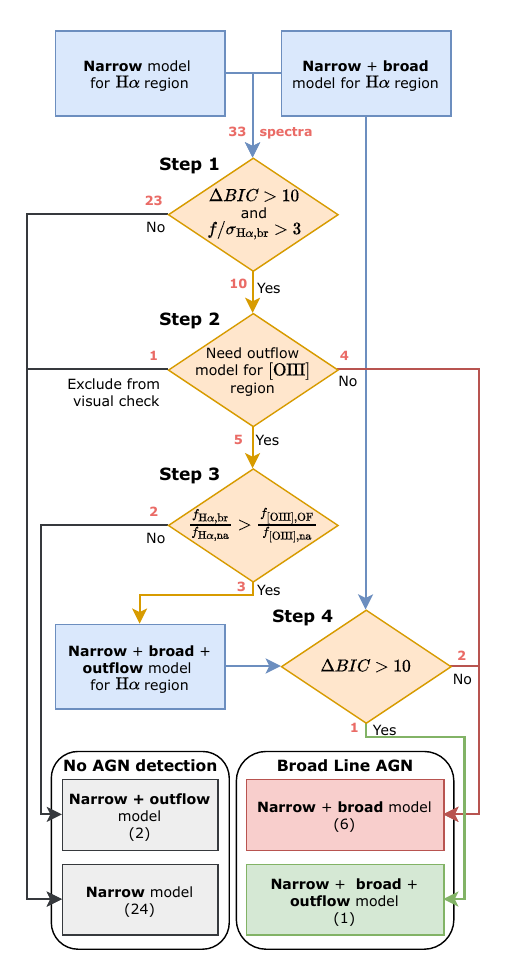}
    \caption{Flowchart of the spectral fitting pipeline for identifying broad-line AGNs. Rectangular boxes describe the model fits to the \Ha\ region with the three final models shown at the bottom. Diamond boxes represent decision criteria for model selection, with red numbers indicating the count of spectra passing or failing each criterion.}
    \label{fig:Flowchart}
\end{figure}

We developed a systematic pipeline to identify and model broad-line AGN signatures in our sample. The workflow of our analysis is illustrated in Fig. \ref{fig:Flowchart}, with the key steps being: (1) testing for the presence of an additional broad \Ha\ component beyond the single narrow-line model, (2) examining the \OIII\ doublets for evidence of outflows, (3) evaluating whether the broad \Ha\ component could originate from outflows rather than an AGN, and (4) attempting to fit a three-component model to the \Ha\ region when appropriate. This systematic approach was applied to all aperture spectra extracted from our IFU data.

We initially focus on the \Ha\ spectral region (rest-frame $6400-6900$ \AA), which contains five strong emission lines: \Ha, \NIIdoublet\ and \SIIdoublet. We first model the local continuum using a power-law function, fitted to three line-free windows ($6400-6500$ \AA, $6650-6700$ \AA, and $6750-6900$ \AA). After continuum subtraction, we fit each emission line with a single Gaussian profile, constraining all lines to share identical line widths and velocity offsets. The flux ratio of the \NIIdoublet\ is fixed to 2.96, which is derived from quantum mechanical calculations of transition probabilities \citep{Osterbrock2006}. We refer to this as our narrow model (top-left box in Fig. \ref{fig:Flowchart}). All fitting is performed using the Levenberg-Marquardt algorithm implemented from the {\tt lmfit} package \citep{MattNewville2024}, with parameter uncertainties derived through Markov chain Monte Carlo (MCMC) sampling.

We then test whether an additional broad component improves the fit. In this narrow$+$broad model (top-right box in Fig. \ref{fig:Flowchart}), we constrain the narrow emission lines to have FWHM $<$ 600 \kms\ and add a broad Gaussian component with FWHM $>$ 600 \kms\ to model potential broad \Ha\ emission. This broad component is allowed to vary freely in width, amplitude, and central wavelength.

We evaluate the necessity of including a broad component by comparing the Bayesian Information Criterion \citep[BIC,][]{Liddle2007} between the narrow and narrow$+$broad models (\textbf{step 1} in Fig. \ref{fig:Flowchart}). The broad component is considered significant if:
\begin{equation}
   \Delta\text{BIC} = \text{BIC}_{\text{narrow}} - \text{BIC}_{\text{narrow+broad}} > 10,
\end{equation}
where the BIC is defined as
\begin{equation}
   \text{BIC} = \sum_{i}{\frac{(O_i-C_i)^2}{\sigma_i^2}} + k\ln N.
\end{equation}
Here, the first term is the $\chi^2$ function where $O_i$ and $C_i$ represent the observed and model fluxes at each wavelength channel, with $\sigma_i$ denoting the flux uncertainty. The term $k\ln N$ penalises model complexity, where $k$ is the number of free parameters and $N$ is the number of data points. Additionally, we require the broad \Ha\ component to be detected with high significance to its uncertainty:
\begin{equation}
   f_\mathrm{H\alpha,br} / \sigma_\mathrm{H\alpha,br} > 3,
\end{equation}
where $f_\mathrm{H\alpha,br}$ and $\sigma_\mathrm{H\alpha,br}$ are the flux and its uncertainty for the broad component.

This procedure identified broad \Ha\ components in 10 of our 33 aperture spectra, with three of these apertures associated with a single merger system (\textit{DC\_848185}). We note that \textit{DC\_842313} lacks G395M coverage and thus sufficient wavelength range for \Ha\ analysis. Its strong outflow signature \citep{Solimano2025} also precludes reliable identification of any broad \Hb\ component in the available G235M data. We therefore exclude this target from subsequent analysis. While our fitting confirms the presence of broad emission components in these spectra, this alone does not definitively establish AGN detections. Outflows can produce similar spectral signatures, necessitating careful consideration of alternative explanations, as discussed in the following section.

\subsection{Modeling Outflows} \label{subsec:outflow}

Outflow features are commonly observed in AGN spectra, particularly in the \OIII\ emission lines \citep[e.g.,][]{Mullaney2013,Woo2016,KovacevicDojcinovic2022}. A typical outflow signature manifests as a blue-shifted, broader component in addition to the core narrow component, usually with velocity dispersions of $200-500$ \kms\ \citep[e.g.,][]{Sexton2021}. Recent JWST observations have revealed similar outflow signatures in high-redshift galaxies, detected in both \Ha\ and \OIII\ emission lines \citep[e.g.,][]{Tang2023b, Xu2023,Carniani2024,Zhang2024c,Bertola2025}. Since outflows can potentially mimic broad-line signatures in the \Ha\ region, we first examine the \OIII\ doublet for outflow features and use the relative strengths of outflow components to evaluate the authenticity of detected broad \Ha\ components.

We analyze the emission lines within the rest-frame spectral window of $4750-5150$~\AA\ (hereafter, the \OIII\ region) using a sequential approach. While this region contains both the \OIIIdoublet\ and \Hb\ lines, we first focus on modeling the \OIII\ doublet independently, as it is typically the strongest feature and the most sensitive tracer of outflows. We model the narrow component of the \OIIIdoublet\ with two Gaussian profiles, fixing their flux ratio to the theoretical value of 2.98 \citep{Storey2000}. The continuum is estimated from line-free windows at $4750-4800$~\AA\ and $5050-5150$~\AA. After completing the \OIII\ analysis, we use these results to guide the \Hb\ line fitting, as detailed below.

To model potential outflows, we perform the fitting on the \OIII\ doublet within the wavelength range of $4925-5050$~\AA. We add a second, broader Gaussian component to each line of the doublet, constraining them to have the same line width and velocity shift relative to the narrow components. We adopt this narrow$+$outflow model only if it improves the fit by $\Delta\text{BIC}>10$ compared to the narrow-component-only model (\textbf{step 2} in Fig. \ref{fig:Flowchart}). This analysis identified outflow signatures in five of our initial ten AGN candidates.

In addition, we identified 1 case (\textit{DC\_873756}) with no detectable emission lines in the \OIII\ region. Visual inspection of the \Ha\ region of this spectrum revealed an unusual profile dominated by broad emissions with minimal narrow-line contribution, contrary to expectations for main-sequence galaxies. Given the low S/N and ambiguous nature of its broad \Ha\ component, we exclude this target from our final AGN candidate list. A detailed analysis of this dismissed case is presented in Appendix \ref{Appendix}.

Once the best-fitting model for the \OIII\ doublet is determined (either with or without an outflow component), we use this line profile as a template to fit the \Hb\ emission. Specifically, we shift the entire \OIII\ model to the expected wavelength of \Hb\ and fit only for its amplitude. This approach is adopted because the \Hb\ line is often too weak to independently constrain complex models with multiple components. By fixing the line shape to that of the much higher S/N \OIII\ profile, we obtain a more robust measurement of the total \Hb\ flux, which is crucial for reliable emission-line diagnostics.

With the outflow properties identified by the \OIII\ doublet, we return to the \Ha\ region to verify the detected broad-line signature of these AGNs. An optimal approach would be to fit the \Ha\ and \OIII\ regions simultaneously to robustly separate the various kinematic components. However, a joint fitting is technically challenging for this dataset. The large wavelength separation often places these line complexes on different NIRSpec gratings (G395M and G235M). Even when on the same grating, instrumental effects such as wavelength calibration uncertainties and the complex, wavelength-dependent line spread function make a simultaneous fit unreliable, potentially creating artificial outflow features \citep{Graaff2024,Maiolino2024}. Therefore, we adopt a more conservative, sequential approach.

Empirically, outflow signatures tend to be more prominent in the \OIII\ doublet than in other emission lines \citep{KovacevicDojcinovic2022}. To avoid misidentifying outflow components as AGN broad-line emission, we require:
\begin{equation} \label{eq:line_ratio_criterion}
   \frac{f_{\rm H\alpha,br}}{f_{\rm H\alpha,na}} > \frac{f_{\rm \OIII,OF}}{f_{\rm \OIII,na}},
\end{equation}
where $f_{\rm H\alpha,br}$ and $f_{\rm H\alpha,na}$ represent the broad and narrow \Ha\ component fluxes, while $f_{\rm \OIII,OF}$ and $f_{\rm \OIII,na}$ denote the outflow and narrow component fluxes in \OIII\ (\textbf{step 3} in Fig. \ref{fig:Flowchart}). This criterion ensures that the broad component detected in \Ha\ is significantly more prominent than any outflow component seen in \OIII, increasing the credibility of the AGN identification. This criterion excludes 2 of our 9 AGN candidates, including 1 from \textit{DC\_848185}. The remaining 7 aperture spectra constitute our final AGN candidates, which we describe in detail in Sec. \ref{sec:BLAGN}.

For candidates where an outflow is detected in \OIII, we cautiously explore a three-component model (narrow$+$broad$+$outflow) for the \Ha\ region. However, such a decomposition is often degenerate. If the black hole mass of these SMBHs follow the local relation, the expected broad-line widths of these low mass AGNs can be comparable to those of outflows, making it difficult to robustly separate the two components based on the \Ha\ profile alone. Applying a strict $\Delta\text{BIC}>10$ criterion for adding the third component, we find that only one candidate (our most robust detection, see Sec. \ref{subsec:most_reliable}) yields a stable and physically plausible three-component solution (\textbf{step 4} in Fig. \ref{fig:Flowchart}). For the other cases, we retain the more conservative two-component (narrow$+$broad) model for \Ha, acknowledging that the broad component parameters may be influenced by some contribution from an underlying outflow.

We note that we do not test the narrow$+$outflow model for the \Ha\ region without a broad component. This is because without clear parameter constraints to distinguish between outflow and broad-line components, the narrow$+$outflow model can mathematically reproduce the narrow$+$broad model. By performing such a fitting, we still cannot confirm nor exclude the existence of AGN broad emission. In addition, we caution that for cases where we do not consider the outflow component (either because the spectra cannot fit a three-component model or the outflow in the \OIII\ region is too weak), the broad \Ha\ component we measured could potentially be partially contaminated by outflow emission. 

Another caveat comes from the ignorance of an outflow contribution from \NIIdoublet. Since the \NII /\Ha\ ratio in our sample is typically low, the outflow contribution from \NIIdoublet~is usually negligible. Nevertheless, it is worth noting that previous studies have shown star formation or shock-driven outflows can sometimes enhance the \NII /\Ha\ ratio \citep{Newman2013,DAgostino2019}. In such instances, the \NII\ outflow component might contribute to the line profile broadening, potentially affecting the derived black hole mass.

\section{Detectability Limits for Broad-Line AGNs} \label{sec:detectability}

Before presenting our results from the ALPINE-CRISTAL-JWST data, it is crucial to understand the limitations in detecting broad-line AGNs. We first conduct simulations to demonstrate the expected emission line profiles of broad-line AGNs with varying Eddington ratios in host galaxies of different masses. We then examine the property ranges of these AGNs that can be identified given our current data quality. Our methodology closely follows previous weak AGN detection studies \citep[e.g.,][]{Harikane2023, Maiolino2024}, differing only slightly in outflow detection criteria. Therefore, our simulations can help illustrate selection biases not only in our work but also in similar JWST studies. We adopt an optimistic scenario to establish the fundamental detection limits for broad-line AGNs at these redshifts.

\subsection{Assumptions and Simulations} \label{subsec:assumptions}

\begin{figure*}
   \includegraphics[width=\textwidth]{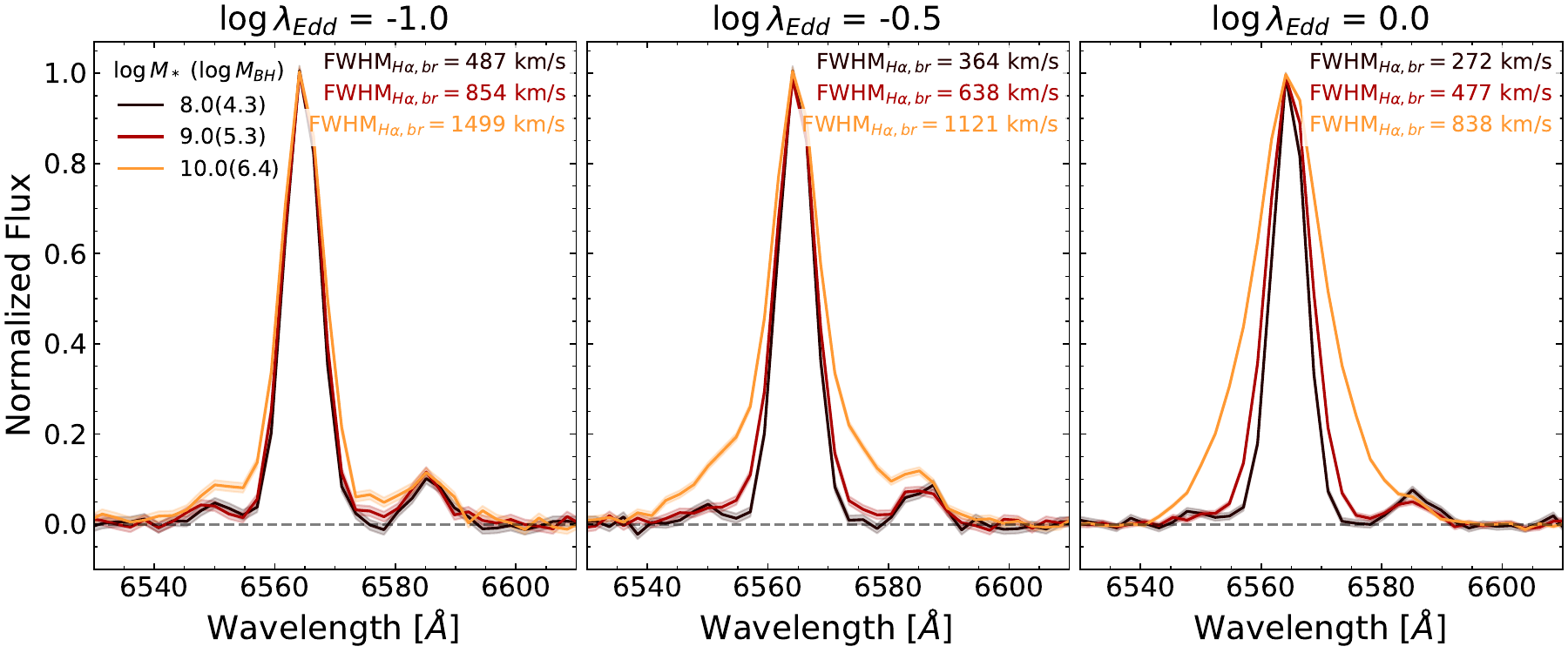}
	\caption{Simulated NIRSpec IFU spectra showing the \Ha\ region for AGNs of varying properties. Each panel shows spectra for three different host galaxy masses ($\log{M_\star/M_\odot} = 8.0, 9.0, 10.0$), with Eddington ratio increasing from left to right ($\log{\lambda_{\rm Edd}} = -1.0, -0.5, 0.0$). Black hole masses and broad \Ha\ line widths are derived using the local scaling relations from \citet{Reines2013,Reines2015}. The spectra include instrumental effects and noise levels matching our ALPINE-CRISTAL-JWST observations.}
	\label{fig:mock_spectra}
\end{figure*}

We begin our detectability assessment by simulating mock \Ha\ region spectra based on local scaling relations and observational conditions. The simulation is parametrized by three quantities: the total stellar mass of the host galaxy ($M_\star$), the Eddington ratio ($\lambda_{\rm Edd}$), and the black hole mass deviation ($\Delta\log{M_{\rm BH}}$) from the local $M_{\rm BH}-M_\star$ relation.

For the mock spectra, we start from the narrow \Ha\ line. Given a stellar mass $M_\star$, we estimate the SFR using the main sequence relation from \citet[][, eq. 28]{Speagle2014}. We assume a redshift of $z=5.0$, based on the redshift distribution of the ALPINE-CRISTAL-JWST sample. From the SFR, we derive the \Ha\ line luminosity using the correlation from \citet[][, eq. 1]{Zhuang2019}, combining the excitation conversion relation from \citet{Kennicutt1998} with the initial mass function from \citet{Kroupa2001}. We adopt fixed line ratios of $\log{f_{\rm [NII]6584}/f_{\rm H\alpha}}=-1.0$ and $\log{f_{\rm [SII]6717,6731}/f_{\rm H\alpha}}=-1.0$, typical values for weak AGNs at high redshift (See Fig. \ref{fig:BPT}). The flux ratio of the \NIIdoublet\ is fixed to its theoretical value of 2.96 (see Section \ref{subsec:fitting}). While the \SIIdoublet\ ratio varies with electron density, we adopt a value of 1.0. This choice has a negligible impact on our results due to the intrinsic weakness of these lines. The FWHM of the narrow lines is set to 350 \kms\ after including the instrumental resolution, which is the median FWHM of the narrow \Ha\ line in our sample. The narrow line width is dominated by the spectral resolution, the small variation between different $M_\star$ is negligible.

Beyond the narrow emission from the host galaxy, we assume only a single Gaussian broad component in \Ha, excluding any systematic velocity shifts or additional outflow components. We also omit potential contributions from the narrow-line region of the AGN. This simplified approach represents an optimistic scenario, as including these additional components would only reduce the detectability of the broad-line signature.

For the $M_{\rm BH}-M_\star$ relation, we adopt the prescription from \citet{Reines2015}. We calculate $M_{\rm BH}$ by combining this relation with the specified $\Delta\log{M_{\rm BH}}$. Assuming a single Gaussian profile, we then determine the luminosity and FWHM of the broad \Ha\ component using $M_{\rm BH}$ and $\lambda_{\rm Edd}$. The following relations are assumed for the solution: 
\begin{enumerate}
   \item the correlation between $L_{\rm H\alpha, br}$ and $L_{5100}$ from \citet[][, eq. 1]{Greene2005}:
   \begin{equation} \label{eq:L_Ha_L5100}
      L_{\rm H\alpha,br} = 5.25 \times 10^{42} \left(\frac{L_{5100}}{10^{44}\ergs}\right)^{1.157};  
   \end{equation}
   \item the bolometric correction at 5100 \AA, ${\rm BC_{5100}}=9.26$ from \citet{Shen2011};
   \item the single-epoch black hole mass estimation recipe from \citet[][, eq. 5]{Reines2013}:
   \begin{equation} \label{eq:M_BH_Reines2013}
      \begin{aligned}
         \log{M_{\rm BH}/M_\odot} = & 6.6 + 0.47\log\left(\frac{L_{\rm H\alpha,br}}{10^{42}\ergs}\right) + \\  
                                    & 2.06\log\left(\frac{\rm FWHM_{\rm H\alpha,br}}{1000\kms}\right),
      \end{aligned}
   \end{equation}
   where $L_{\rm H\alpha,br}$ and FWHM$_{\rm H\alpha,br}$ are the luminosity and FWHM of the broad \Ha\ component, respectively.
\end{enumerate}

To match observational conditions, we incorporate instrumental and data processing effects into our mock spectra. Our IFU aperture spectra are extracted from only the central pixels to maximise sensitivity to point-like AGN emission while minimizing contamination from extended star formation. We adopt a representative value of 15\% for the fraction of total line flux captured in these central apertures, based on the median of our ALPINE-CRISTAL-JWST sample. This fraction varies individually with the changes of galaxy compactness and star formation distribution, ranging from 10\% to 30\% in our sample. We then convolve the spectra to the NIRSpec IFU resolution ($R=1000$) and add uniform random noise to achieve a peak signal-to-noise ratio of 3 in the \NIIleft\ line, matching our observational design.

Fig. \ref{fig:mock_spectra} presents the simulated \Ha\ line profiles for different combinations of $M_\star$ and $\lambda_{\rm Edd}$, processed to match NIRSpec IFU observations. Following the local $M_{\rm BH}-M_\star$ relation from \citet{Reines2015}, the broad component FWHM increases with higher $M_\star$ but decreases with higher $\lambda_{\rm Edd}$. The corresponding $M_{\rm BH}$ and FWHM$_{\rm H\alpha,br}$ values are indicated for each case.

\subsection{Detection of the broad \Ha\ line} \label{subsec:detection}

\begin{figure}
   \includegraphics[width=\columnwidth]{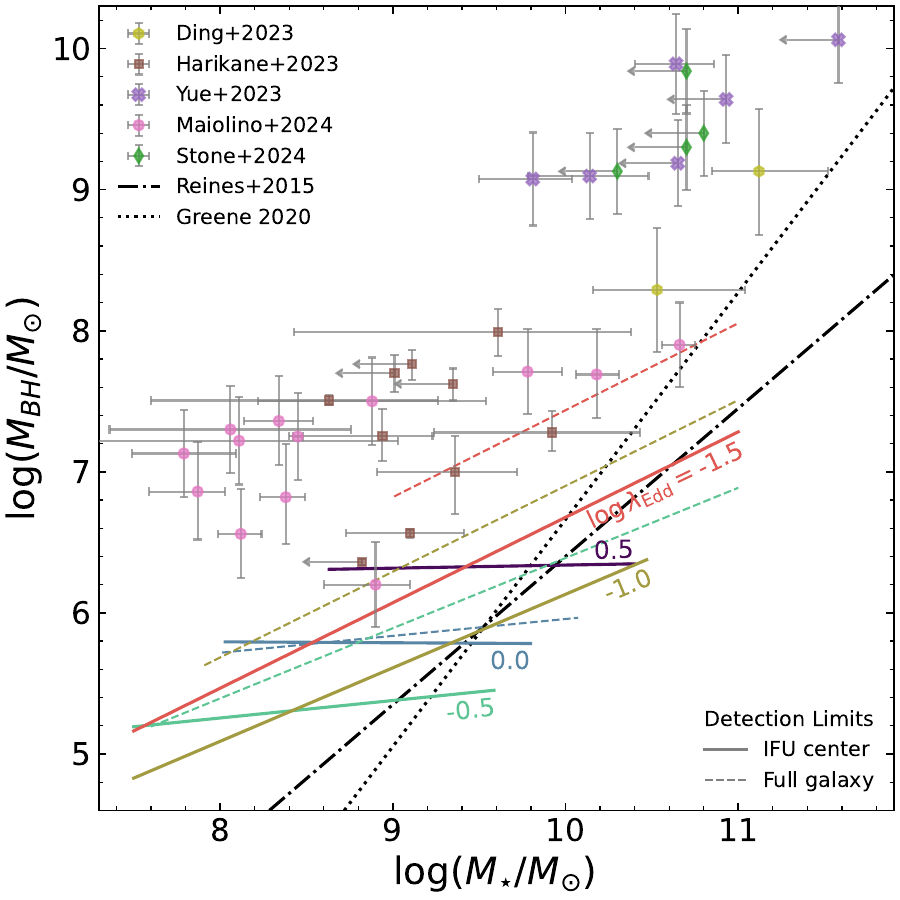}
	\caption{Detection limits for broad-line AGNs in the $M_\star$-$M_{\rm BH}$ plane. Solid and dashed lines show the 50\% detection probability thresholds for different Eddington ratios ($\log{\lambda_{\rm Edd}}$), using central aperture (IFU centre) and integrated galaxy spectra (Full galaxy), respectively. Note that for $\log{\lambda_{\rm Edd}}=0.5$, the `Full galaxy' line overlaps with the `IFU center' line. Black dash-dotted and dotted lines represent local $M_{\rm BH}-M_\star$ relations from \citet{Reines2015} and \citet{Greene2020}. AGN systems below the local relation are unlikely to be detected, especially in low-mass galaxies. Our IFU aperture spectra method provides less biased detection due to reduced contamination from extended galaxy emission. High-redshift AGN detections from recent JWST studies \citep{Ding2023,Harikane2023,Maiolino2024,Stone2024a,Yue2024b} are shown for comparison. All previously detected AGNs lie well above our predicted detection limits and have Eddington ratios of $\log{\lambda_{\rm Edd}}\sim-0.5$, consistent with our most favorable detection scenario.}
	\label{fig:detectability}
\end{figure}

Fig. \ref{fig:mock_spectra} demonstrates that, assuming high-redshift AGNs follow local scaling relations, broad \Ha\ line detection becomes increasingly challenging for lower-mass host galaxies. Here we quantify these detection limits by applying our criteria systematically across the $M_\star$-$M_{\rm BH}$ plane.

We systematically explore a grid of parameters: stellar mass ($7.5 \leq \log{M_\star/M_\odot} \leq 11.0$ in steps of 0.1), black hole mass deviation ($-0.5 \leq \Delta\log{M_{\rm BH}} \leq 1.5$ in steps of 0.1), and Eddington ratio ($-1.5 \leq \log{\lambda_{\rm Edd}} \leq 0.5$ in steps of 0.5). At each grid point, we generate 100 mock spectra and apply our detection criteria. The detection rate is calculated as the fraction of spectra where a broad component is successfully identified.

As illustrated in Fig. \ref{fig:detectability}, the detection probability shows distinct patterns across different Eddington ratio regimes. The solid lines represent the 50\% detection probability thresholds achieved using our central aperture method, while the dashed lines show results from integrated galaxy spectra, highlighting the enhanced sensitivity of the central aperture method. These two methods are intended to bracket the most optimistic (central aperture) and pessimistic (integrated spectrum) scenarios for \Ha\ detection. We note that the sensitivity of Multi-Object Spectrograph (MOS) observations, such as those performed with the NIRSpec MSA, would likely fall between these two extremes. A precise one-to-one comparison is complex, as the true detection limit for a MOS observation depends critically on factors such as the centring of the shutter on the galaxy nucleus and the resulting slit losses, as well as the specific line-fitting methodology employed.

For sub-Eddington accreting AGNs ($\log{\lambda_{\rm Edd}} < 0$), the 50\% detection threshold lines run from the upper-right to lower-left, creating a boundary that separates the detectable AGNs (above/left of the line) from the undetectable ones (below/right of the line). This orientation indicates a detection bias favoring the upper-left region of the parameter space (high $M_{\rm BH}$, low $M_\star$) compared to the lower-right region (low $M_{\rm BH}$, high $M_\star$). Conversely, for super-Eddington accretion ($\log{\lambda_{\rm Edd}} > 0$), the detection threshold lines become more horizontal, indicating that detectability becomes less dependent on host galaxy mass and more strongly limited by the black hole mass itself. This change in behavior is primarily due to the reduced broad line width at high Eddington ratios.

For typical Eddington ratios, our detection threshold lies significantly above the local $M_{\rm BH}-M_\star$ relation for host galaxies with $M_\star < 10^{9.5}M_\odot$. Only in more massive galaxies ($M_\star > 10^{9.5}M_\odot$) can we detect AGNs with black hole masses that follow or fall below the local relation, and even then only for moderate Eddington ratios. This selection bias becomes more pronounced when using integrated galaxy spectra due to increased contamination from narrow emission lines. The fact that all previously detected weak AGNs \citep{Harikane2023,Maiolino2024} fall within our calculated detection limits provides strong support for the validity of our simulation approach.

The detectability of broad-line AGNs is strongly influenced by the Eddington ratio, with optimal detection occurring around $\log{\lambda_{\rm Edd}}\simeq-0.5$. As illustrated in Fig. \ref{fig:mock_spectra}, both very high and very low Eddington ratios present distinct challenges for detection. At high Eddington ratios, the broad \Ha\ line width decreases significantly, making it difficult to distinguish from narrow emission components. We stress that this is a methodological limit, imposed to ensure a robust separation between the broad and narrow line components, rather than a fundamental limit imposed by signal-to-noise. Despite adopting an aggressive threshold of ${\rm FWHM}_{\rm H\alpha,br} > 600$ \kms\ in our pipeline, which is less conservative than in other studies, we remain unable to detect AGNs with $M_{\rm BH}<10^{5.8}M_\odot$ accreting at the Eddington limit. This provides an optimistic reference for broad-line AGN detectability. Since the broad line width depends primarily on $M_{\rm BH}$ at fixed $\lambda_{\rm Edd}$, this creates an approximately horizontal detection threshold in the $M_\star$-$M_{\rm BH}$ plane that rises with increasing $\lambda_{\rm Edd}$. Conversely, while lower Eddington ratios produce broader lines that are more easily distinguished from narrow components, they also result in weaker emission. The combination of large line width and low luminosity means the peak flux of the broad component can fall below our significance requirement. Consequently, detection sensitivity also decreases at low Eddington ratios, particularly for less massive black holes where the emission is inherently weaker.

Our simulations demonstrate that selection effects can bias the detection of broad-line AGNs towards higher black hole masses, potentially contributing to the apparent prevalence of overmassive black holes in high-redshift galaxies. To improve sample completeness in future studies, we recommend several strategies. 
First, investigations of scaling relations at high redshift should be aware of the detection challenges of lower-mass black holes. AGNs hosted by galaxies with $M_\star \gtrsim 10^{9}M_\odot$ are more viable to be detected assuming a local $M_{\rm BH}-M_\star$ relation. Second, high spectral resolution is crucial, particularly for galaxies with $M_\star \lesssim 10^{10}M_\odot$, where broad \Ha\ line widths typically do not exceed 1000 \kms\ at moderate Eddington ratios. Finally, spatially resolved IFU observations can significantly improve detection sensitivity by minimizing contamination from extended star formation.

Several caveats should be considered when interpreting our simulation results. First, our assumption of a single Gaussian broad-line profile may be oversimplified. In reality, the presence of both intermediate and very broad components could enhance detectability \citep[e.g.,][]{Sulentic2000a}. Second, while continuum subtraction is negligible for our ALPINE-CRISTAL-JWST sample, it could significantly impact broad-line detection in cases where the continuum is stronger. Finally, although our simple spectral model enables reliable property measurements when broad components are detected, the measurements on real observations may show systematic biases depending on data quality and line profile complexity.

\section{Characterization of Broad-Line AGN Candidates} \label{sec:BLAGN}

Here, we present our 7 broad-line AGN candidates from the ALPINE-CRISTAL-JWST sample and detail the justification for each identification. Two candidates are hosted by a single merger galaxy system (\textit{DC\_848185}). These candidates were selected through our systematic spectral analysis pipeline, which identified significant broad \Ha\ components that satisfy our detection criteria outlined in Section \ref{subsec:fitting}.

Throughout this paper, we derive the AGN bolometric luminosities and virial black hole masses following equations \ref{eq:L_Ha_L5100} and \ref{eq:M_BH_Reines2013} from \citet{Greene2005} and \citet{Reines2013}, with a bolometric correction of ${\rm BC_{5100}}=9.26$ from \citet{Shen2011}. Line fluxes are corrected for dust attenuation using the Balmer decrement (\Ha/\Hb\ ratio of the narrow component), adopting the attenuation curve from \citet{Calzetti2000}. For host galaxy stellar masses, we adopt the values from \citet{Faisst2020}, which were derived through broad-band photometric SED fitting. Table \ref{tab:AGN_prop} summarizes the measured AGN properties.

\subsection{Robust Identification: \textit{DC\_536534}} \label{subsec:most_reliable}

\begin{figure*}
   \includegraphics[width=\textwidth]{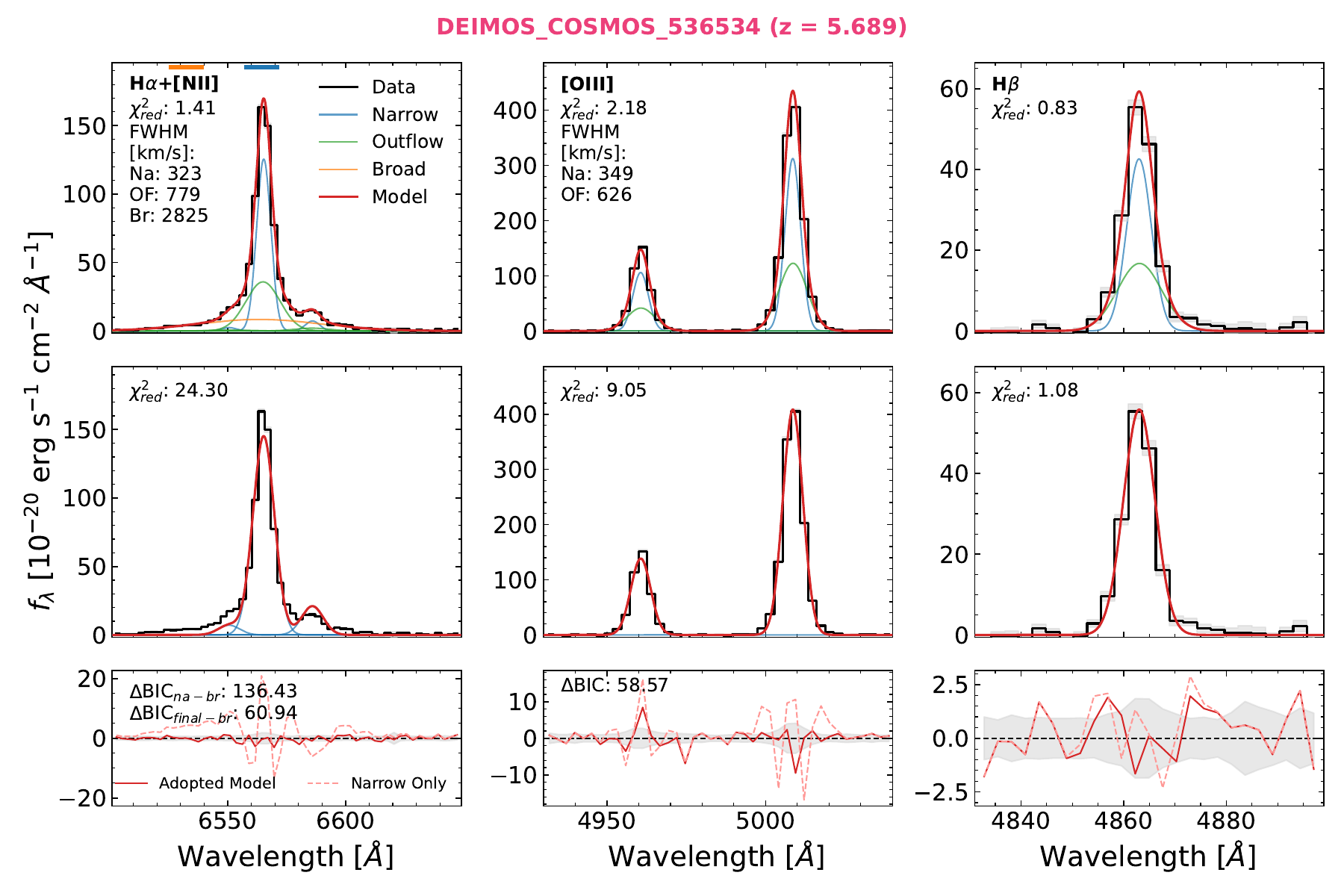}
   \caption{Spectral analysis of our most robust AGN candidate, \textit{DC\_536534} (HZ1/CRISTAL-03). Panels show fits to the \Ha$+$\NII\ (left), \Hb\ (middle), and \OIII\ (right) emission lines. Top row: fits incorporating narrow$+$broad$+$outflow components; middle row: fits using only narrow components. Reduced $\chi^2$ and measured FWHM values are indicated for each fit (note that these FWHM values are not corrected for instrumental broadening). Bottom panels show fit residuals, demonstrating the necessity of the broad \Ha\ component. We show the $\Delta$BIC on the top left. The $\Delta$BIC$_{na-br}$ compares the BIC value of narrow model and narrow$+$broad model for \Ha~and the $\Delta$BIC$_{na-br}$ tests between narrow$+$broad and narrow$+$broad$+$outflow models. The instrumental dispersion corrected FWHM of broad \Ha\ is provided in Table \ref{tab:AGN_prop}. Blue and orange horizontal line segments in the top left panel mark wavelength ranges used for spatial analysis in Fig. \ref{fig:QA4HabMap}.}
   \label{fig:QA4Convincing}
\end{figure*}

\begin{figure}
   \includegraphics[width=\columnwidth]{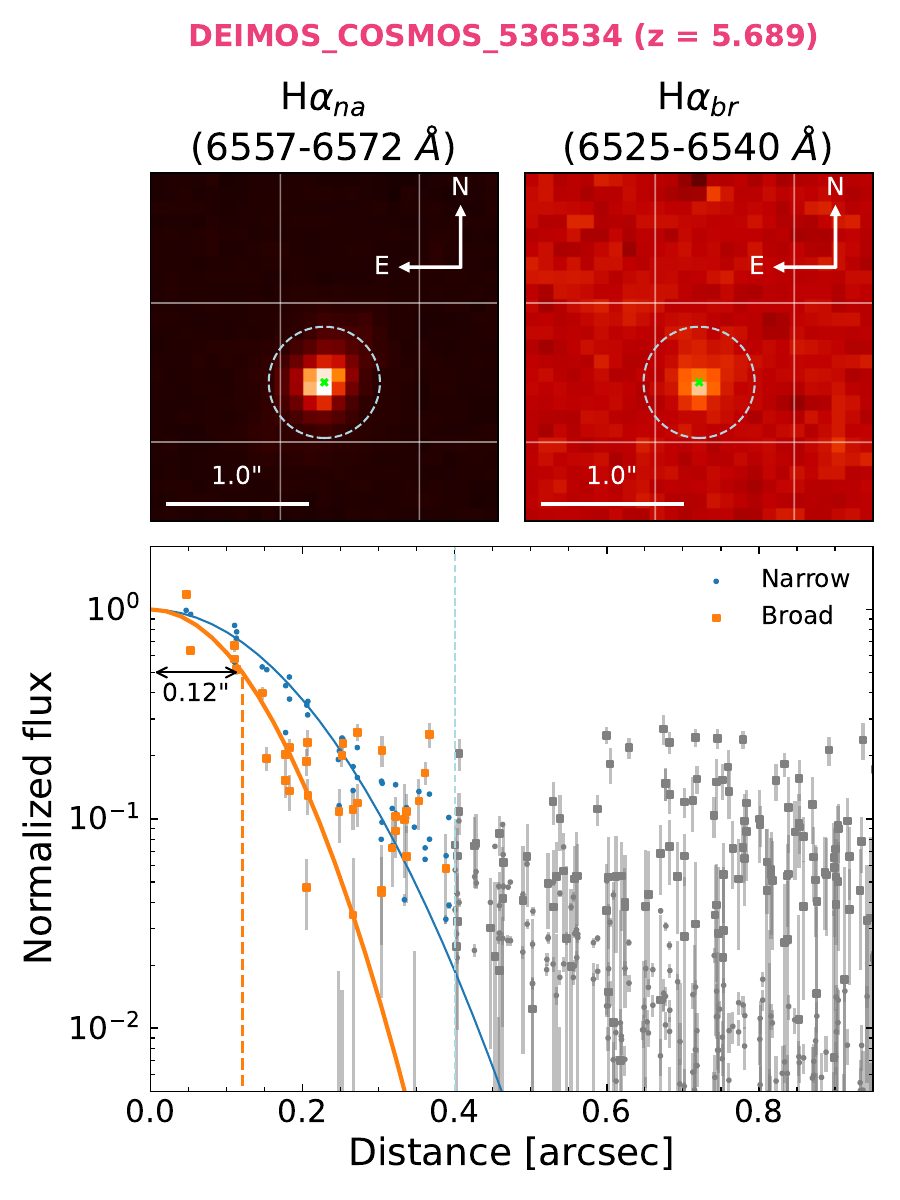}
   \caption{Spatial analysis of the narrow and broad \Ha\ components in \textit{DC\_536534}. Upper panels show mean flux maps for two wavelength ranges: $6557-6572$ \AA\ (left, narrow \Ha) and $6525-6540$ \AA\ (right, broad \Ha\ blue wing). Gaussian profiles are fitted to pixels within 0.4\arcsec\ radius of the centre (light blue circles). The lower panel presents normalised flux of these two components versus radial distance, with best-fitting Gaussian models. The smaller FWHM of the broad \Ha\ blue wing (0.24$\pm$0.01\arcsec\ versus 0.33$\pm$0.01\arcsec\ for narrow) confirms its more compact spatial distribution, consistent with an AGN origin.}
   \label{fig:QA4HabMap}
\end{figure}

We identify \textit{DC\_536534} (also known as HZ1 and CRISTAL-03) as our most robust AGN candidate. Fig. \ref{fig:QA4Convincing} presents the spectral fitting for the \Ha\ region using a narrow$+$broad$+$outflow model, alongside fits for the \OIII\ region using an outflow$+$narrow model. For comparison, fits of narrow model are shown in the second row. The inclusion of the broad component significantly reduces the $\chi^2$ of the fit.

The fitted broad component in the \Ha\ region exhibits a FWHM of $\sim$2800 \kms, consistent with typical AGN BLR. Its corresponding flux is $f_{\rm H\alpha,br}=5.6\pm0.7\times 10^{-18}~\ergscm$. We designate this target as our most robust AGN candidate based on its characteristic broad \Ha\ line width. This target is unique in our sample for having a broad \Ha\ component FWHM exceeding 2000 \kms, a width commonly observed in local AGNs and notably higher than typical outflow velocities \citep{Mullaney2013}. In addition to the broad \Ha\ line, we successfully isolated the outflow component, which is also detected independently in the \OIII\ lines, thus strengthening the interpretation of the broad \Ha\ line as originating from the BLR.

Taking advantange of the high FWHM of broad \Ha\ line (and its extension towards the blue side of the \NIIleft\ line), we use our IFU data to examine the spatial distribution of this broad component. Fig. \ref{fig:QA4HabMap} presents two mean flux maps covering different rest-frame wavelength ranges: $6557-6572$ \AA\ (encompassing the narrow \Ha\ line) and $6525-6540$ \AA\ (capturing the blue wing of the broad \Ha\ line). These wavelength ranges are indicated in the top left panel of Fig. \ref{fig:QA4Convincing}. We subtract the continuum flux from each pixel, estimated using the same continuum windows employed in our \Ha\ region fitting. In the lower panel of Fig. \ref{fig:QA4HabMap}, we present the 1D profile of line flux versus distance from our aperture centre. The flux decreases rapidly from the centre and becomes dominated by noise beyond 0.4\arcsec. Gaussian profile fits to both flux maps reveal a distinct difference in FWHM between the narrow and broad components. The broad component exhibits a FWHM of 0.24$\pm$0.01\arcsec, closer to the PSF FWHM of NIRSpec, confirming that this emission originates from an unresolved, point-like source as expected for the BLR. In contrast, the narrow component shows a broader FWHM of 0.33$\pm$0.01\arcsec, indicating it is partially resolved and includes contributions from the extended host galaxy. This difference in spatial extent aligns with expectations: the broad component originates from the central AGN's BLR (unresolved at NIRSpec resolution), while the narrow component includes contributions from the more extended host galaxy.

Our choice of a narrow$+$broad$+$outflow model for the \Ha\ region of \textit{DC\_536534} warrants further explanation. Among our sample, this target uniquely requires three Gaussian components to adequately model its \Ha\ line profile. We interpret the intermediate-width component as outflow-driven emission and the broadest component as originating from the BLR, using only the latter for AGN property calculations. Several factors support this decomposition:
\begin{enumerate}
    \item Detection of an outflow component with similar width in both \Hb\ and \OIIIdoublet\ lines;
    \item Matching FWHM between the \Ha\ and \OIII\ outflow components;
    \item The \Ha\ outflow-to-narrow flux ratio remains below the corresponding ratio for \OIII.
\end{enumerate}
These characteristics provide strong evidence for identifying the intermediate-width component as outflow-driven emission. Notably, combining the outflow and BLR components into a single broad line would significantly reduce the measured FWHM, thereby biasing the inferred $M_{\rm BH}$ towards lower values (see Fig. \ref{fig:MM_relation}).

\subsection{Six other AGN Candidates} \label{subsec:DC417567}

\begin{figure*}
   \centering
   \includegraphics[width=\textwidth]{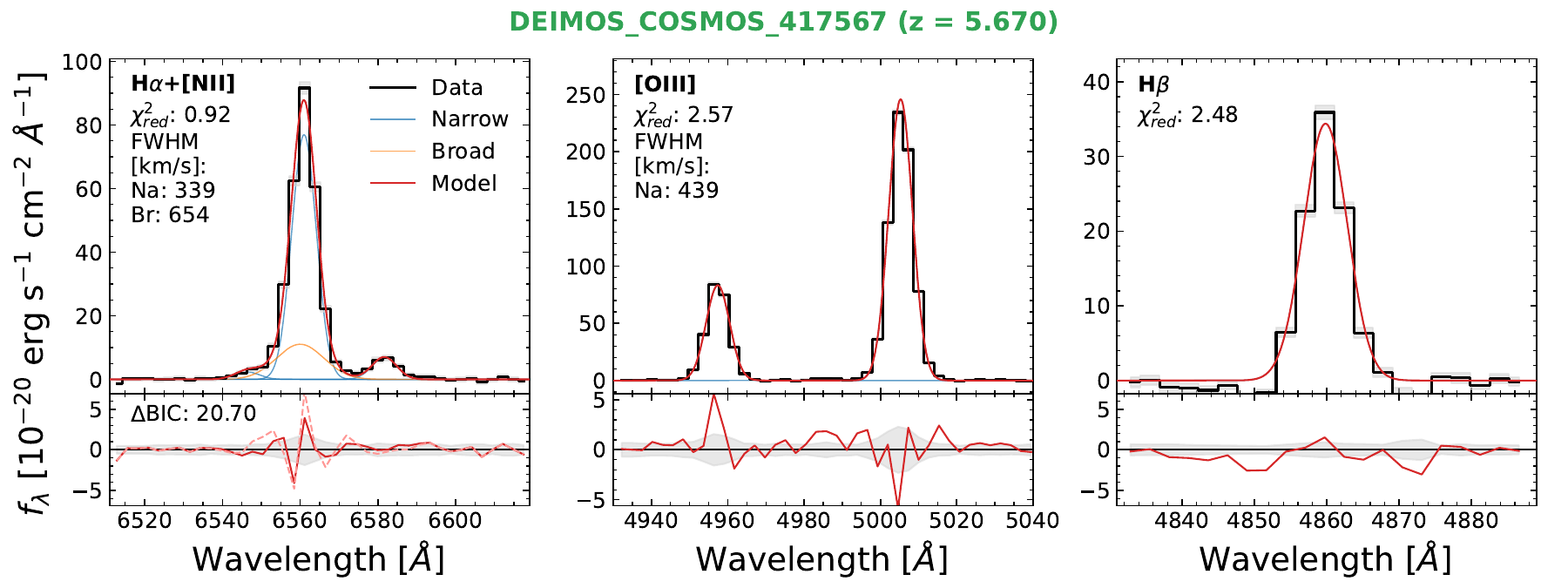}
   \includegraphics[width=\textwidth]{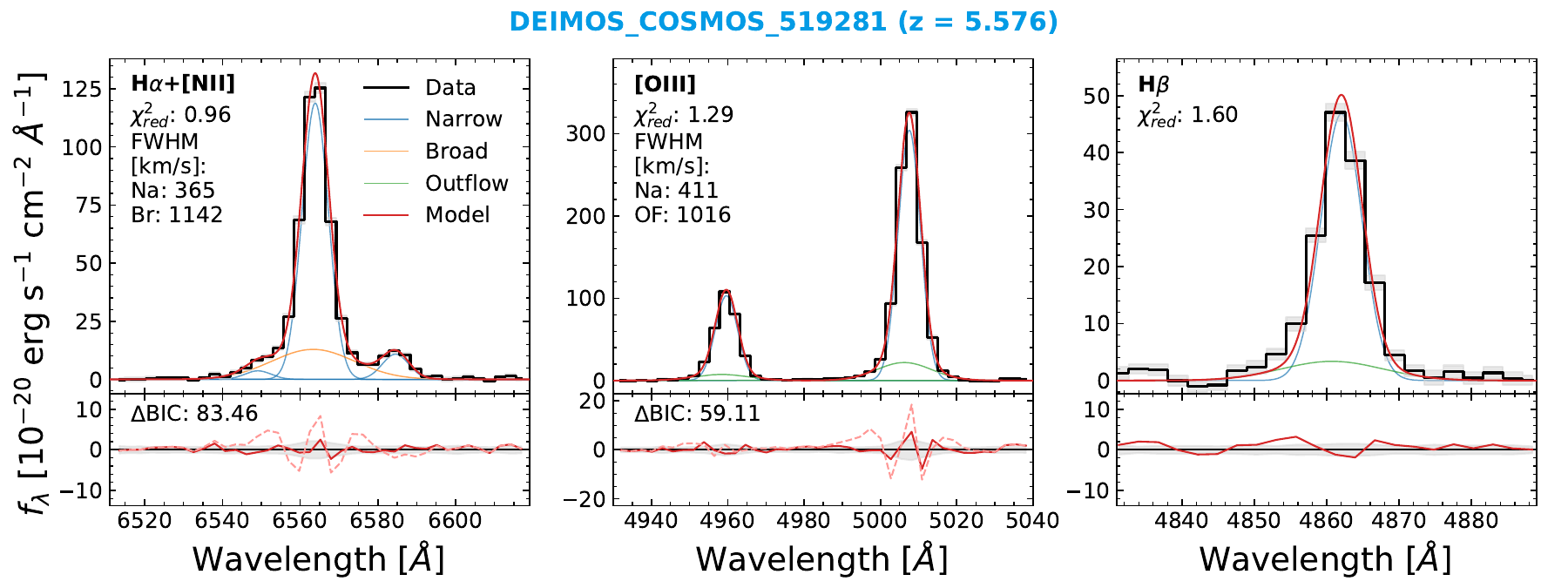}
   \includegraphics[width=\textwidth]{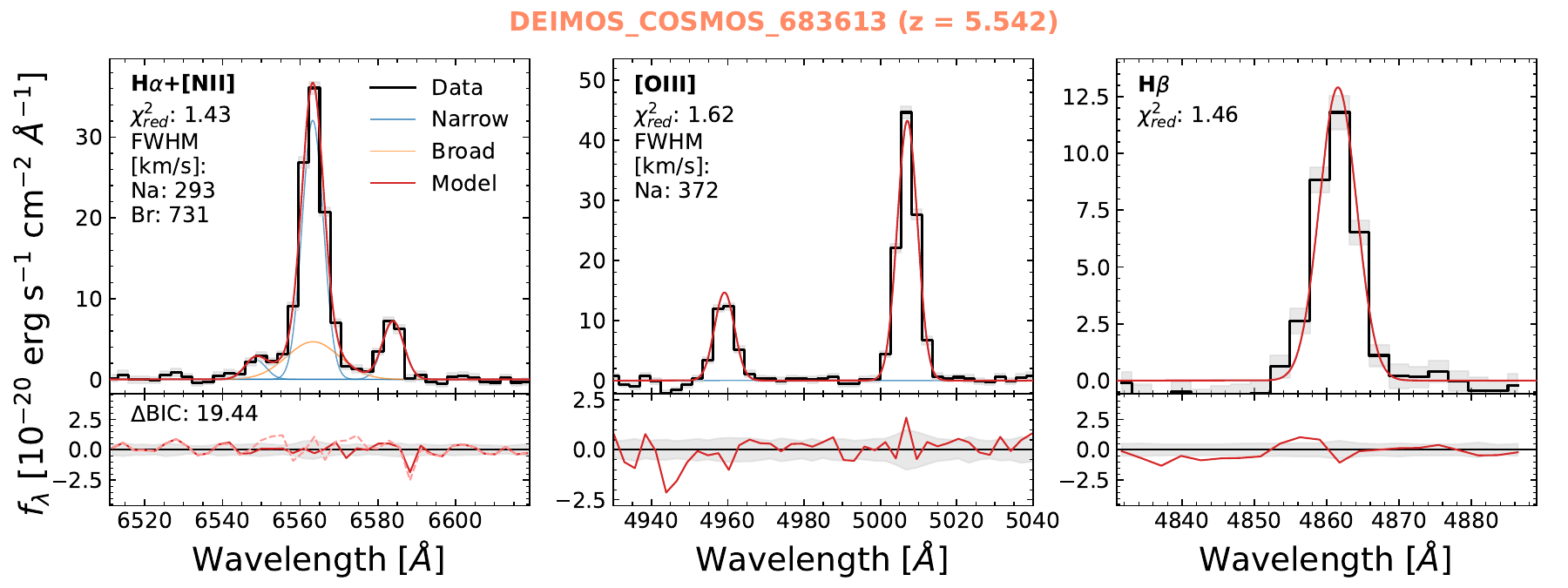}
   \caption{Spectral analysis of three AGN candidates: \textit{DC\_417567} (top), \textit{DC\_683613} (middle), and \textit{DC\_519281} (bottom). Each row shows fits to the \Ha$+$\NII, \Hb, and \OIII\ emission lines. Reduced $\chi^2$ and measured FWHM values are indicated for each fit (note that the FWHM values are not corrected for instrumental broadening). Bottom panels display model fit residuals (solid lines), with comparisons to narrow-only model residuals (dashed lines) and the $\Delta$BIC if applicable. Measurement uncertainties are shown as shaded regions. Target names are color-coded for consistent identification throughout the paper.}
   \label{fig:QA4AGN_1}
\end{figure*}

\begin{figure*}
   \centering
   \includegraphics[width=\textwidth]{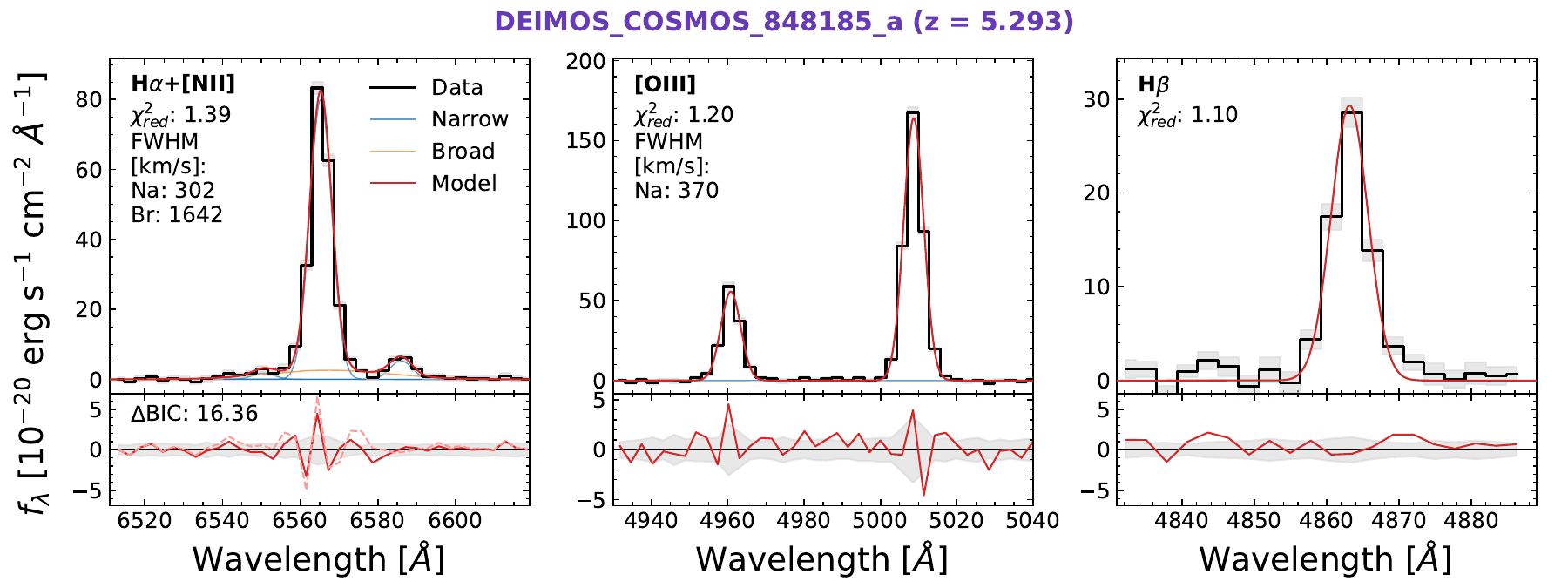}
   \includegraphics[width=\textwidth]{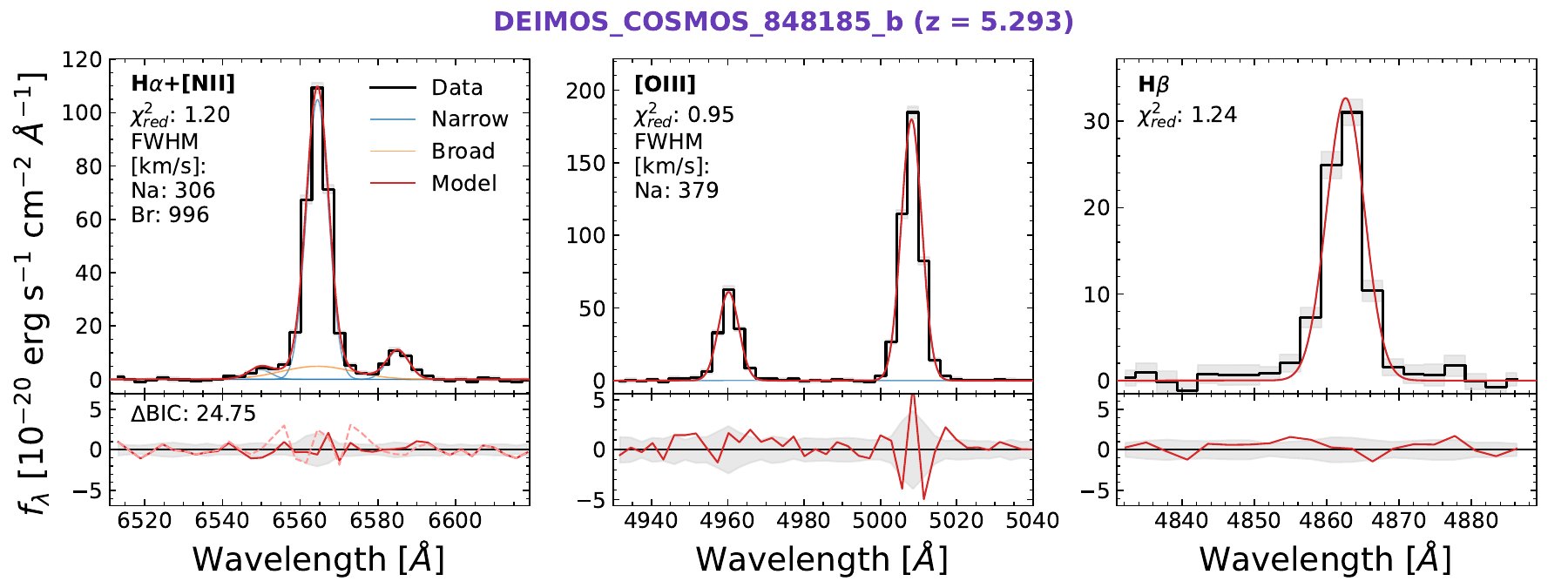}
   \includegraphics[width=\textwidth]{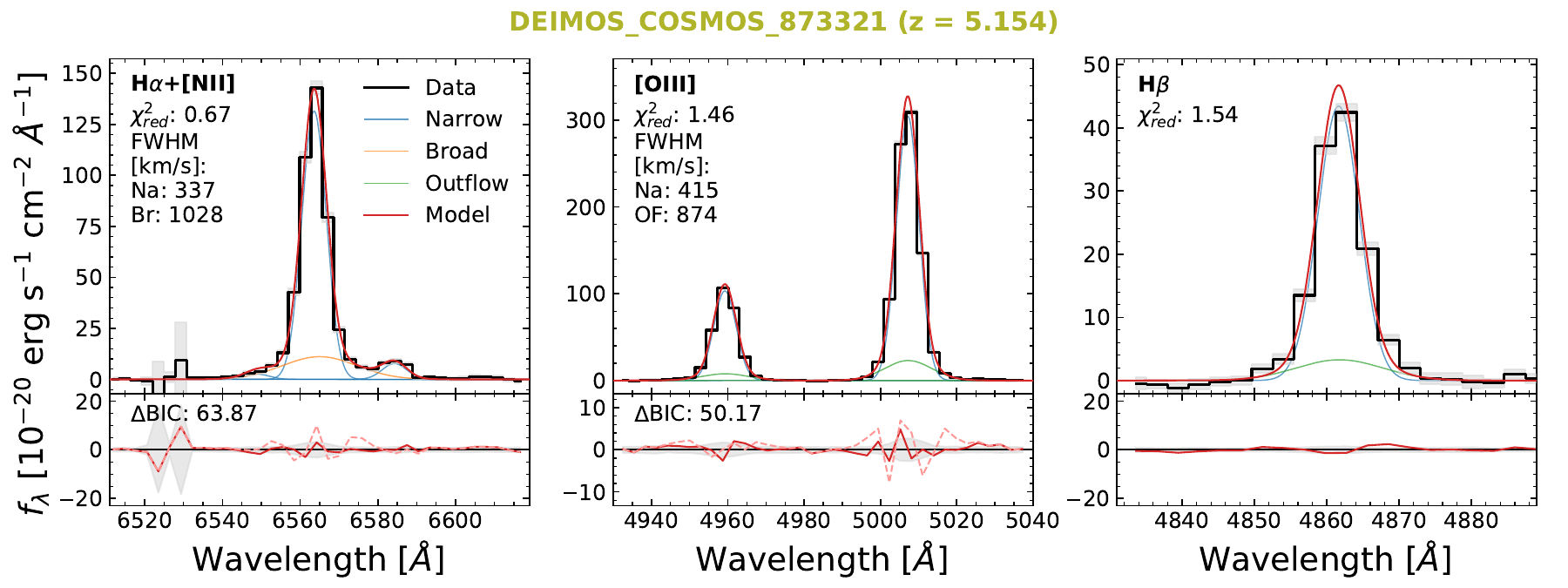}
   \caption{Spectral analysis of additional AGN candidates: \textit{DC\_848185} peaks \texttt{a} and \texttt{b} (top and middle panels), and \textit{DC\_873321} (bottom). Panel layout and markings follow the format of Fig. \ref{fig:QA4AGN_1}. The FWHM values shown are not corrected for instrumental broadening.}
   \label{fig:QA4AGN_2}
\end{figure*}

The other six targets differ from \textit{DC\_536534} by exhibiting broad \Ha\ components with FWHM ranging from 600 to 1600 \kms. Unlike our most robust case, these moderate broad line widths cause the \Ha\ to significantly blend with the \NIIdoublet, preventing us from isolating a clean broad component wavelength region for spatial distribution analysis as we did for \textit{DC\_536534}. Moreover, their FWHM values fall within a range that could be produced by outflows rather than necessarily indicating BLR emission, making their AGN interpretation less definitive. Nevertheless, as we described in Sec. \ref{subsec:outflow}, we use the \OIII\ line as a probe of possible outflow components. Figs. \ref{fig:QA4AGN_1} and \ref{fig:QA4AGN_2} present our spectral analysis of these candidates. Here, we describe each one separately.

\textbf{\textit{DC\_417567}} (HZ2/CRISTAL-10) shows a broad \Ha\ component with a relatively small width of \mbox{FWHM$_{\rm corr}$ = 596 \kms} after instrumental dispersion correction. The broad component flux, $f_{\rm H\alpha,br}=1.7\pm0.4\times 10^{-18}~\ergscm$, barely exceeds 4$\sigma$ significance. While the broad \Ha\ evidence is modest, the narrow line ratios place this source in the AGN region of both \NII\ and \SII\ BPT diagrams \citep{Baldwin1981}. Importantly, we detect no outflow signatures in the \OIII\ region.

\textbf{\textit{DC\_519281}} (CRISTAL-09) shows a broad \Ha\ component with \mbox{FWHM$_{\rm corr}$ = 1110 \kms} and $f_{\rm H\alpha,br}=3.4\pm0.3\times 10^{-18}~\ergscm$. We detect an outflow component in \OIII\ with a modest systematic velocity offset of $\sim$80 \kms. Although the outflow component width matches the broad \Ha\ component, the outflow-to-narrow flux ratio in \OIII\ ($f_{\rm [OIII],OF}/f_{\rm [OIII],na}=0.18\pm0.03$) is substantially lower than that of broad \Ha\ ($f_{\rm H\alpha,br}/f_{\rm H\alpha,na}=0.34\pm0.03$). The broad and narrow \Ha\ components share identical central wavelengths. The different flux ratios and velocity characteristics suggest the broad \Ha\ emission likely originates from the BLR rather than the outflow detected in \OIII.

\textbf{\textit{DC\_683613}} (HZ3/CRISTAL-05) exhibits a broad \Ha\ component with \mbox{FWHM$_{\rm corr}$ = 678 \kms} and $f_{\rm H\alpha,br}=0.80\pm0.14\times 10^{-18}~\ergscm$. The fitting to the \OIII\ region rejects an outflow scenario. We note that this target has a relatively strong \NII\ and \SII\ line flux with $\log{(f_{\rm \NII}/f_{\rm H\alpha})}=-0.64\pm0.04$ and $\log{(f_{\rm \SII}/f_{\rm H\alpha})}=-0.88\pm0.06$, significantly higher than the rest of our sample (see Sec. \ref{subsec:BPT}). With a relatively lower $\log{(f_{\rm [OIII]}/f_{\rm H\alpha})}=0.54\pm0.02$, this target situates in the local composite region of the BPT diagram \citep{Baldwin1981}. We note that it is confirmed as a merger system according to its \CII\ observation \citep{Posses2024} and has an off-centre dust geometry traced by far-infrared observation \citep{Killi2024}. Disfavoring the outflow cause of the dust geometry, this observation further indicates the possibility of AGN feedback in this system.

\textbf{\textit{DC\_848185}} (HZ6/LBG-1/CRISTAL-02) presents a complex merger system with two AGN candidates. Our \Ha\ line flux map reveals four photometric peaks labelled as \texttt{a-d} in the lower left panel of Fig. \ref{fig:AGN_morph}. The peak \texttt{a} is relatively isolated from the other three kinematically connected peaks \texttt{b-d}. Our analysis identifies AGN candidates in two locations (peaks \texttt{a} and \texttt{b}). The northwestern candidate (peak \texttt{a}) exhibits a broad \Ha\ component with \mbox{FWHM$_{\rm corr}$ = 1618 \kms} and $f_{\rm H\alpha,br}=1.01\pm0.16\times 10^{-18}~\ergscm$, with no detected outflow signatures in \OIII.

The southeastern candidate (peak \texttt{b}) shows a broad \Ha\ component with \mbox{FWHM$_{\rm corr}$ = 955 \kms} and $f_{\rm H\alpha,br}=1.14\pm0.18\times 10^{-18}~\ergscm$. While \CII\ observations reveal strong outflows in this region \citep[][in prep]{Davies2025}, the outflow is centred at peak \texttt{d}. Our analysis confirms \OIII\ outflow signatures at peak \texttt{d} but finds no outflow evidence at peak \texttt{b}. The negligible velocity offset between broad and narrow \Ha\ components ($\Delta v_{\rm br-na}=-1\pm42~\kms$) further supports a non-outflow origin for the broad emission at this location.

\textbf{\textit{DC\_873321}} (HZ8/CRISTAL-07) exhibits a broad \Ha\ component with \mbox{FWHM$_{\rm corr}$ = 986 \kms} and $f_{\rm H\alpha,br}=2.67\pm0.4\times 10^{-18}~\ergscm$. While this component shows a slight blueshift ($\Delta v_{\rm br-na}=-65\pm43~\kms$), the detected \OIII\ outflow displays distinctly different characteristics, with a smaller velocity offset ($\Delta v_{\rm OF-na}=-6\pm14~\kms$) and narrower line width. These differing kinematic signatures suggest separate physical origins for the broad \Ha\ and \OIII\ outflow components.

\section{Results} \label{sec:results}

\subsection{$M_{\rm BH}-M_\star$ relation} \label{subsec:MBH_Mstar}

\begin{figure*}
   \includegraphics[width=\textwidth]{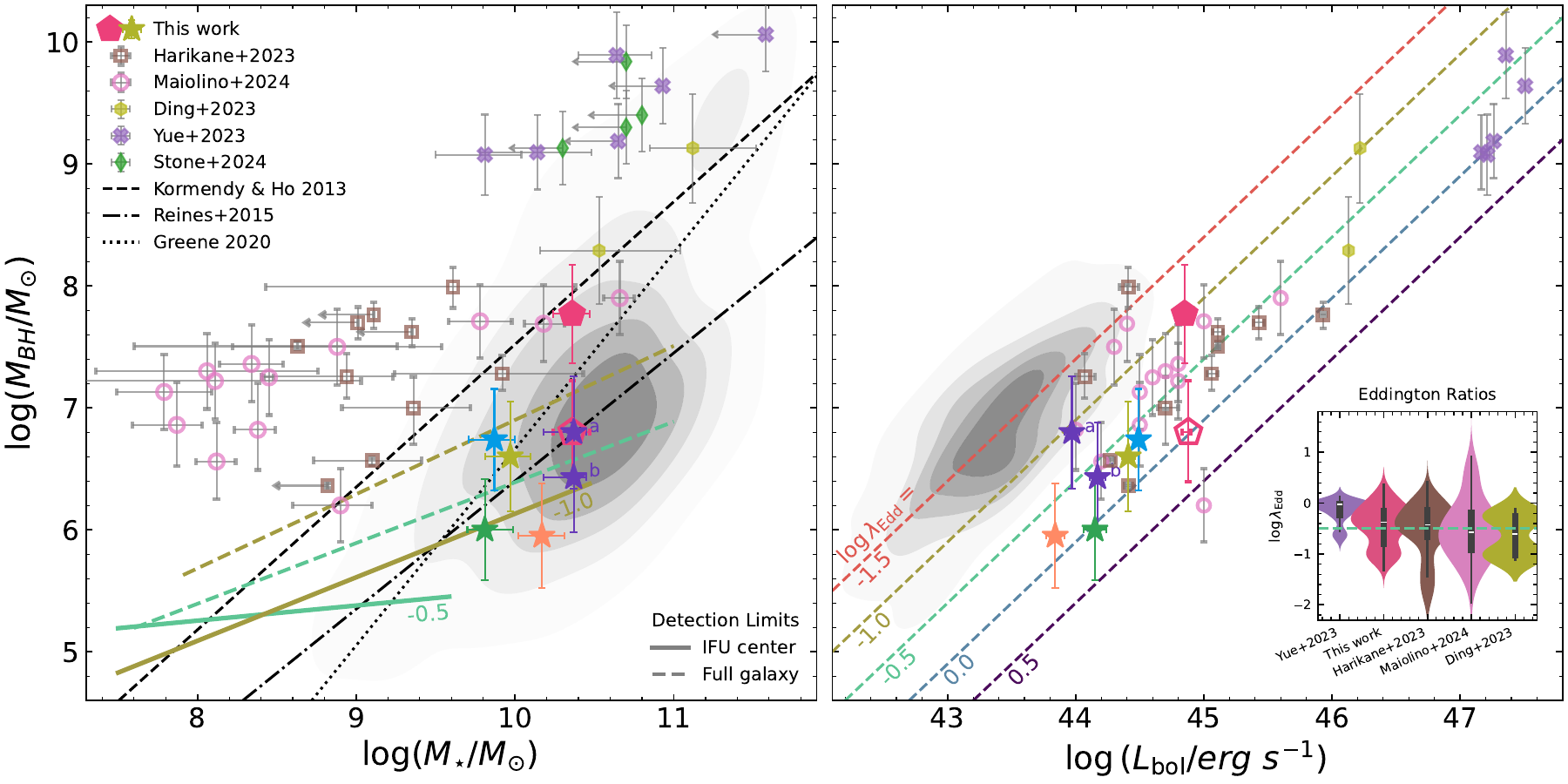}
   \caption{$M_{\rm BH}-M_\star$ relation (left) and the $M_{\rm BH}-L_{\rm bol}$ relation (right) for our AGN candidates compared to other samples. Our robust AGN candidate is shown as a solid pentagon, with a hollow pentagon indicating the $M_{\rm BH}$ estimate when including the outflow component. Other candidates are shown as coloured stars. For comparison, we include high-redshift AGN samples from recent JWST studies, distinguishing between faint AGNs (hollow markers; \citealt{Harikane2023,Maiolino2024}) and more luminous quasars (filled markers; \citealt{Ding2023,Stone2024a,Yue2024b}). Grey contours show the distribution of local AGNs from \citet{Reines2015}. For the left panel, we also include detection limits at $\log{\lambda_{\rm Edd}}=-0.5$ and $-1.0$ as estimated in Sec. \ref{sec:detectability}. Local scaling relations are shown as: the $M_{\rm BH}-M_{\rm bulge}$ relation (dashed line; \citealt{Kormendy2013}) and two $M_{\rm BH}-M_\star$ relations derived from low-redshift AGNs (dotted and dash-dotted lines; \citealt{Reines2015,Greene2020}). The right panel includes an inset violin plot showing the Eddington ratio distribution for each sample. The sample from \citet{Stone2024a} is not included in the right panel due to the absence of reported bolometric luminosities.}
   \label{fig:MM_relation}
\end{figure*}

Studies of the black hole scaling relationship with host galaxy properties in low-z universe have revealed a tight correlation between the black hole mass $M_{\rm BH}$ and the host galaxy stellar mass $M_\star$, with little or no evolution across $z\lesssim2$ \citep[e.g.,][]{Kormendy2013,Ding2020,Tanaka2025}. However, recent JWST observations of high-redshift broad-line AGNs generally report significantly overmassive black holes relative to these local relations \citep{Harikane2023,Maiolino2024,Stone2024a,Yue2024b}. While selection effects likely contribute substantially to this apparent evolution (See Sec. \ref{sec:detectability}; \citealt{Li2024}), the possibility of intrinsic evolution in these scaling relations remains debated. In Fig. \ref{fig:MM_relation}, we show the distribution of our AGN candidates on the $M_{\rm BH}-M_\star$ plane.

The $M_{\rm BH}$ estimates for our candidates are derived from their broad \Ha\ components, as detailed in Sec. \ref{sec:BLAGN}. This single-epoch virial black hole mass measurement technique, while widely used, is subject to a series of systematic uncertainties. Therefore, beyond the statistical uncertainties from our measurements, we incorporate an additional 0.4 dex systematic uncertainty in our $M_{\rm BH}$ estimates \citep{Vestergaard2006,Reines2013}. For \textit{DC\_536534}, where we detect both broad and outflow components in \Ha, we plot both the $M_{\rm BH}$ derived from the broadest component alone (solid pentagon) and from the combination of broad and outflow components (hollow pentagon). The $M_\star$ values are taken from the integrated SED fitting analysis of \citet{Faisst2020}, who utilized broad-band photometry spanning $0.3-8.0$ $\mu$m and derived stellar masses using the {\tt LePhare} code \citep{Ilbert2006}. Our AGN candidates span a black hole mass range of $10^6$ to $10^{7.5}$ $M_{\odot}$, hosted by galaxies with stellar masses from $10^{9.5}$ to $10^{10.5}$ $M_{\odot}$, representing the detection of relatively lower-mass black holes in high-mass galaxies than previously reported at these redshifts. We have verified that our conclusions remain robust when using alternative $M_\star$ estimates from the CRISTAL team \citep{Mitsuhashi2024}.

Unlike many previous JWST studies that have reported a population of `overmassive' black holes at high redshift, our AGN candidates lie either close to or below the local scaling relations \citep{Kormendy2013,Reines2015,Greene2020}. We attribute this finding to two key aspects of our study design. First, the ALPINE-CRISTAL-JWST sample specifically targets representative main-sequence galaxies with $M_\star>10^{9.5}~M_{\odot}$, avoiding the bias towards low-mass galaxies where AGN detection becomes increasingly challenging. Second, our observational strategy combines consistent NIRSpec signal-to-noise ratio (designed to achieve S/N$>$3 in \NII) with spatially resolved IFU data, enabling uniform broad-line detection across the sample while minimizing host galaxy contamination. These factors allow us to identify AGN signatures with reduced selection bias (see Sec. \ref{sec:detectability} for detailed simulations).

For comparison, we examine five other high-redshift AGN samples from JWST observations \citep{Ding2023,Harikane2023,Maiolino2024,Stone2024a,Yue2024b}. The luminous quasar samples ($M_{\rm BH}>10^9~M_{\odot}$) are inherently biased towards more massive black holes. As these are typically from dedicated follow up of pre-selected candidates, they focus on the most luminous, readily detected sources. This selection approach naturally populates the high-mass end of the distribution and can systematically shift the observed $M_{\rm BH}$ above the intrinsic $M_{\rm BH}-M_\star$ relation by approximately an order of magnitude, particularly for high-mass galaxies \citep{Li2022,Yue2024b}.

The samples of fainter broad-line AGNs face different selection challenges. These studies \citep{Harikane2023,Maiolino2024}, which report finding AGNs with $M_{\rm BH}\sim10^7~M_{\odot}$, rely solely on broad-line detection in blind galaxy surveys. As demonstrated in Fig. \ref{fig:detectability}, the intrinsically narrow ``broad emission line'' widths of AGNs with $M_{\rm BH}\lesssim10^6~M_{\odot}$ make their detection extremely challenging at these redshifts with current instrumental capabilities. Given that low-mass galaxies dominate the JWST Advanced Deep Extragalactic Survey (JADES) and Cosmic Evolution Early universe Release Science Survey (CEERS) samples \citep{GomezGuijarro2023,Simmonds2024}, it is likely that these observational limitations have caused surveys to miss a significant population of lower-mass black holes that follow standard relations. This possibility should be considered when interpreting the properties of the detected AGN samples.

Moreover, the detection of AGNs with $M_{\rm BH}<10^6~M_{\odot}$ (commonly referred to as intermediate-mass black holes) remains challenging even in the local universe \citep{Greene2007,Dong2012,Reines2013,Baldassare2016,Mezcua2017,Liu2018,Lin2024}. Beyond the observational difficulties posed by their low luminosities and narrow emission lines, there remains ongoing debate about whether low-mass AGNs can maintain broad-line regions at all \citep{Elitzur2009,Chakravorty2013,Cann2019}. As black hole mass decreases, the temperature of accretion disc and physical conditions in the immediate vicinity of the black hole may change significantly, potentially altering the structure and properties of the broad-line region. Consequently, both technical limitations and physical considerations necessitate caution when applying current broad-line AGN identification methods in this low-mass regime.

Our survey is strategically designed to circumvent the aforementioned selection biases, thereby filling a critical gap in the observed parameter space. By targeting galaxies with $M_\star > 10^{9.5}~M_{\odot}$, we can detect black holes that follow the local scaling relations. The fact that our detected AGN population aligns more closely with the local relation than published JWST samples, in conjunction with the observed high AGN fraction (see Section \ref{subsec:AGN_Fraction}), suggests that the $M_{\rm BH}-M_\star$ relation has not undergone significant evolution from $z\sim0$ to $z\sim4-6$.

\subsection{$M_{\rm BH}-L_{\rm bol}$ relation} \label{subsec:MBH_Lbol}

We present the $M_{\rm BH}-L_{\rm bol}$ relation for our AGN candidates in the right panel of Fig. \ref{fig:MM_relation}, alongside high-redshift AGN samples from recent studies \citep{Ding2023,Harikane2023,Maiolino2024,Yue2024b} and low-redshift AGNs from SDSS \citep{Reines2015}. For luminous quasars from \citet{Ding2023} and \citet{Yue2024b}, bolometric luminosities are derived from continuum luminosity at 5100 \AA. For other AGNs, including our candidates, $L_{\rm bol}$ values are calculated from broad \Ha\ luminosity, as the continuum is predominantly contributed by host galaxy emission.

We note that we recalculate the $L_{\rm bol}$ values for AGNs in \citet{Harikane2023} with the broad \Ha\ luminosity only, differing from their original approached which incorporates the total \Ha\ luminosity. The substantial narrow \Ha\ component from star-forming regions in their method increases the estimated $L_{\rm bol}$ by approximately $0.4$ dex, placing their AGNs near the Eddington limit.

An intriguing pattern emerges in the right panel of Fig. \ref{fig:MM_relation}, where high-redshift luminous quasars, high-redshift weak AGNs, and SDSS low-redshift AGNs occupy three distinct accretion regimes. Despite spanning a range of two orders of magnitude in $M_{\rm BH}$, the Eddington ratios of weak AGNs discovered by JWST distribute around $\log{\lambda_{\rm Edd}}=-0.5$, precisely the value our simulations (Sec. \ref{sec:detectability}) identify as optimal for broad \Ha\ detection. High-redshift luminous quasars exhibit higher $\lambda_{\rm Edd}$ values approaching the Eddington limit, due to different selection biases \citep{Lauer2007}. Both high-redshift populations accrete at significantly higher Eddington ratios than the low-redshift population.

A phase of rapid growth of AGNs at high-redshift is expected, as the gas-rich environment of host galaxies in the early universe facilitates more efficient gas inflow and fueling, leading to a higher accretion rate \citep{Alexander2012}. However, the alignment of our AGN candidates with $\log{\lambda_{\rm Edd}}=-0.5$ likely reflects, at least partly, observational selection effects. The selection approach inherently misses low-$\lambda_{\rm Edd}$ AGNs due to their weak broad \Ha\ emission, and high-$\lambda_{\rm Edd}$ AGNs due to their intrinsically narrower broad emission lines. Consequently, our sample is biased towards AGNs with the most detectable Eddington ratio of approximately $\log{\lambda_{\rm Edd}}=-0.5$. This convergence towards $\log{\lambda_{\rm Edd}} \approx -0.5$ is also seen in other JWST-selected AGN samples at similar redshifts \citep{Harikane2023,Maiolino2024}, suggesting a common selection bias. While the specific line-fitting procedures and sample selection strategies differ between studies, the fundamental sensitivity and spectral resolution of the instrument likely impose a similar detection boundary, favouring this particular accretion regime.

\subsection{Morphology of AGN host galaxies} \label{subsec:morphology}

\begin{figure*}
   \centering
   \begin{minipage}[b]{0.40\textwidth}
      \includegraphics[width=\textwidth]{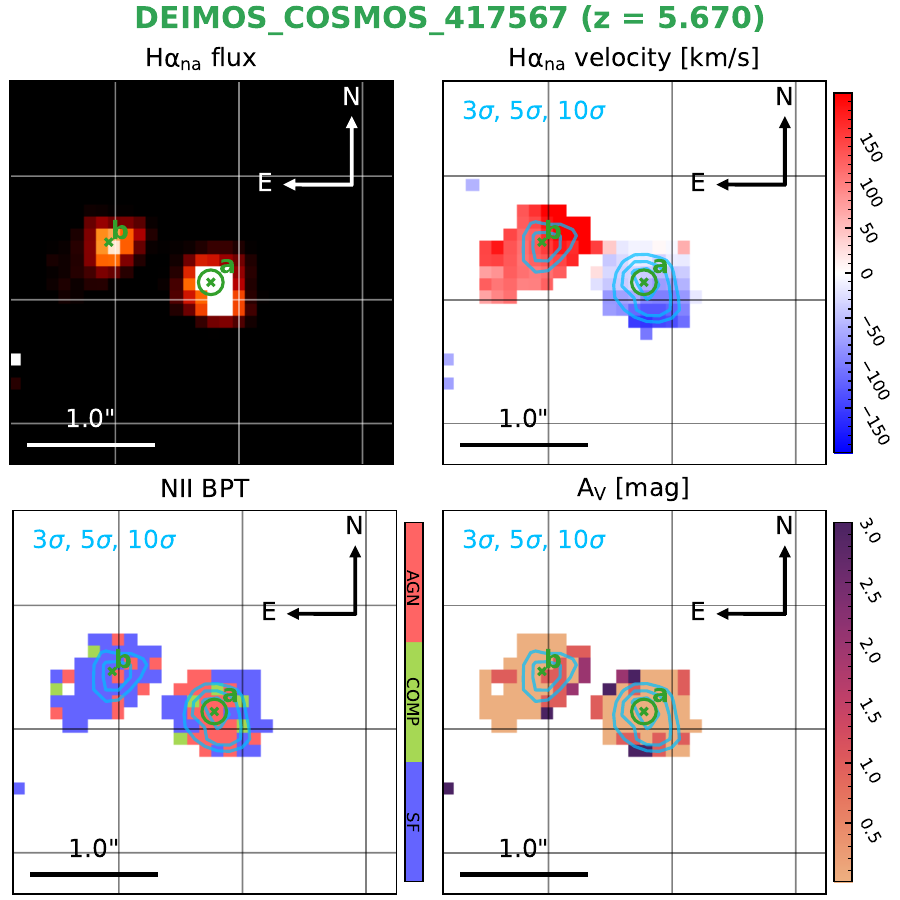}
   \end{minipage}
   \hspace{0.05\textwidth}
   \begin{minipage}[b]{0.40\textwidth}
      \includegraphics[width=\textwidth]{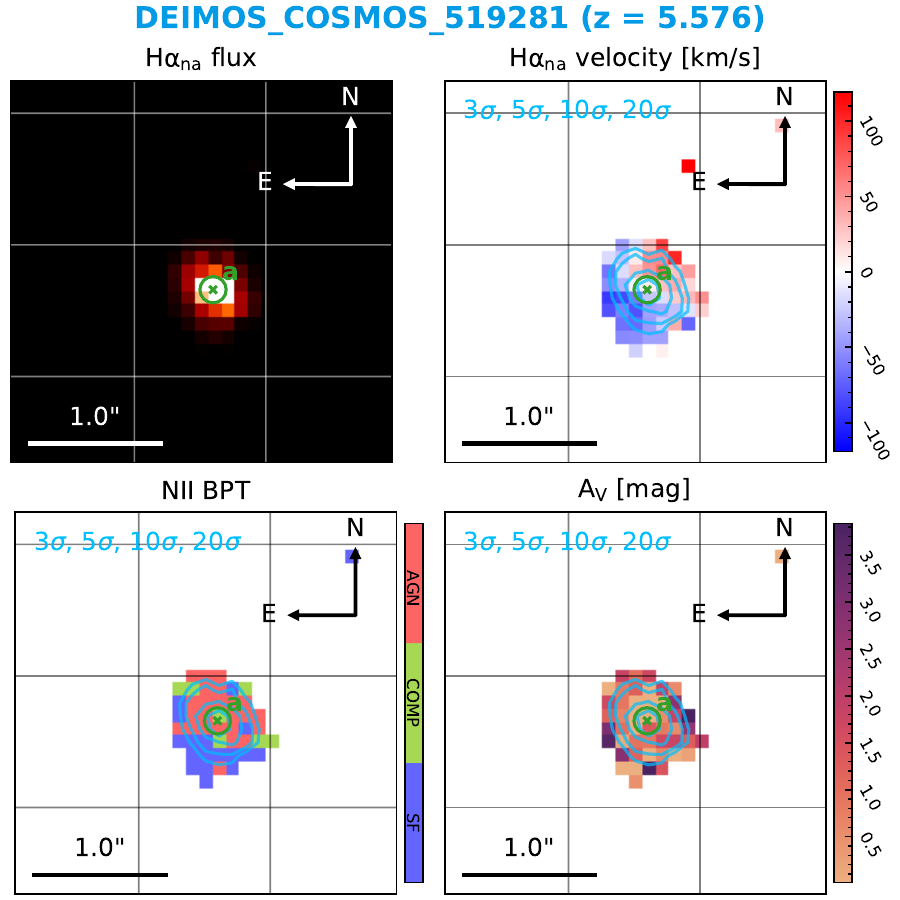}
   \end{minipage}

   \vspace{0.2cm}

   \begin{minipage}[b]{0.40\textwidth}
      \includegraphics[width=\textwidth]{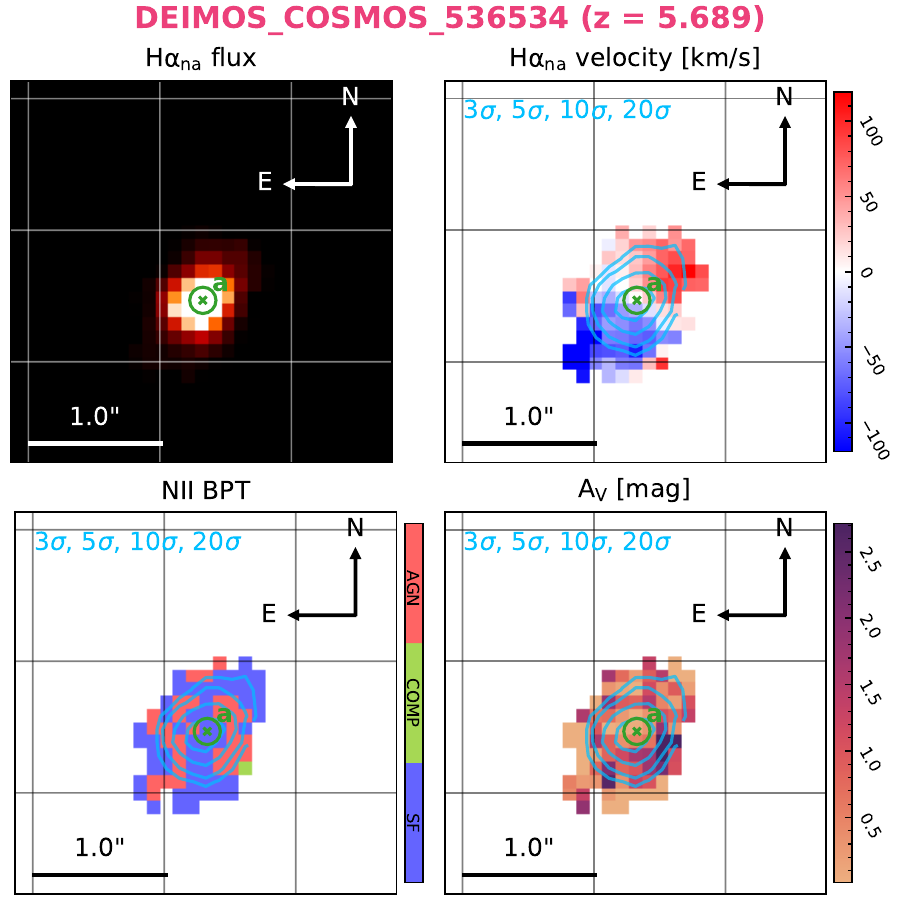}
   \end{minipage}
   \hspace{0.05\textwidth}
   \begin{minipage}[b]{0.40\textwidth}
      \includegraphics[width=\textwidth]{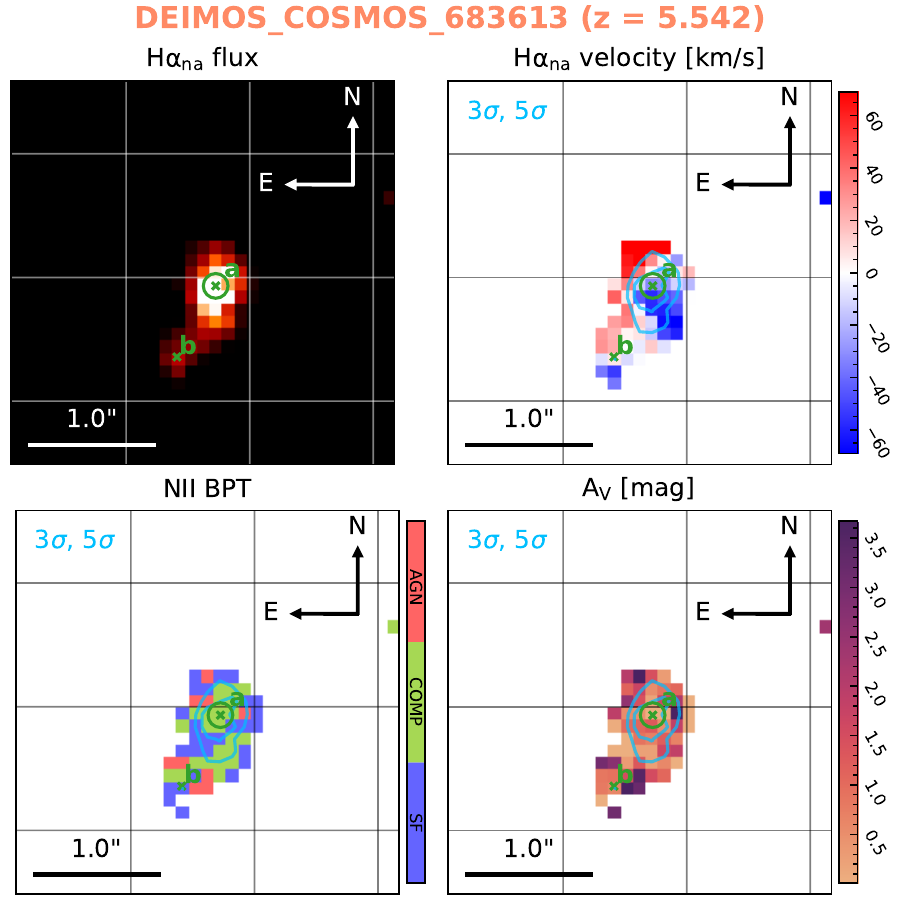}
   \end{minipage}

   \vspace{0.2cm}

   \begin{minipage}[b]{0.40\textwidth}
      \includegraphics[width=\textwidth]{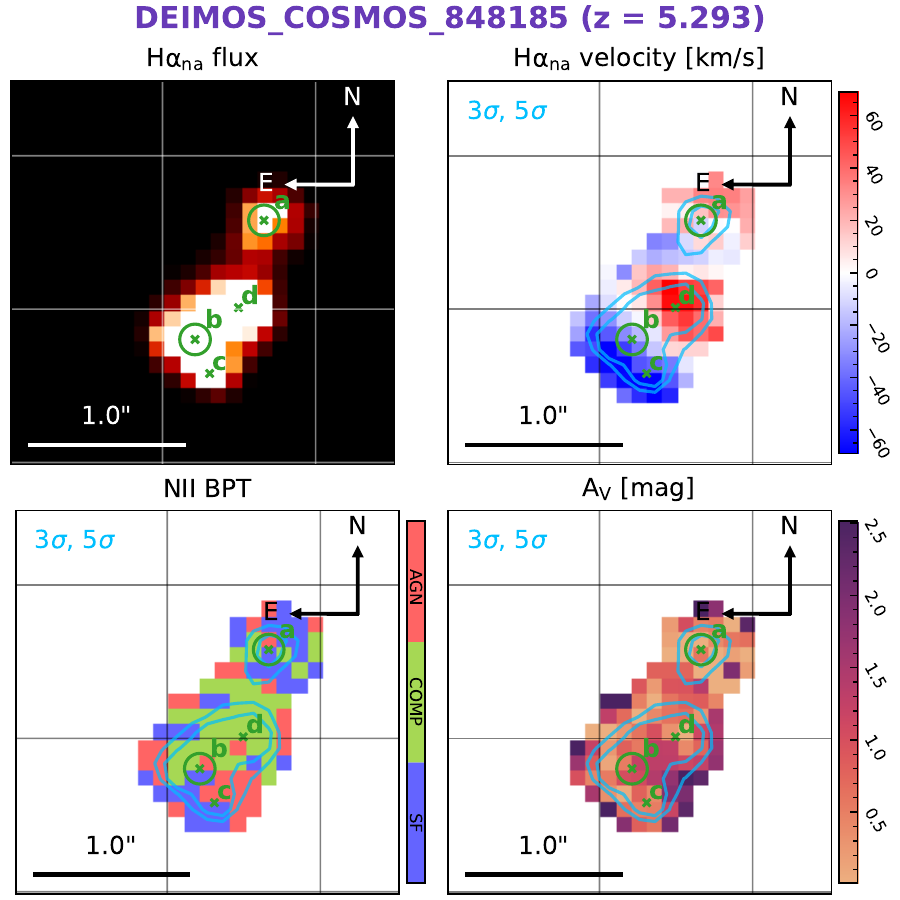}
   \end{minipage}
   \hspace{0.05\textwidth}
   \begin{minipage}[b]{0.40\textwidth}
      \includegraphics[width=\textwidth]{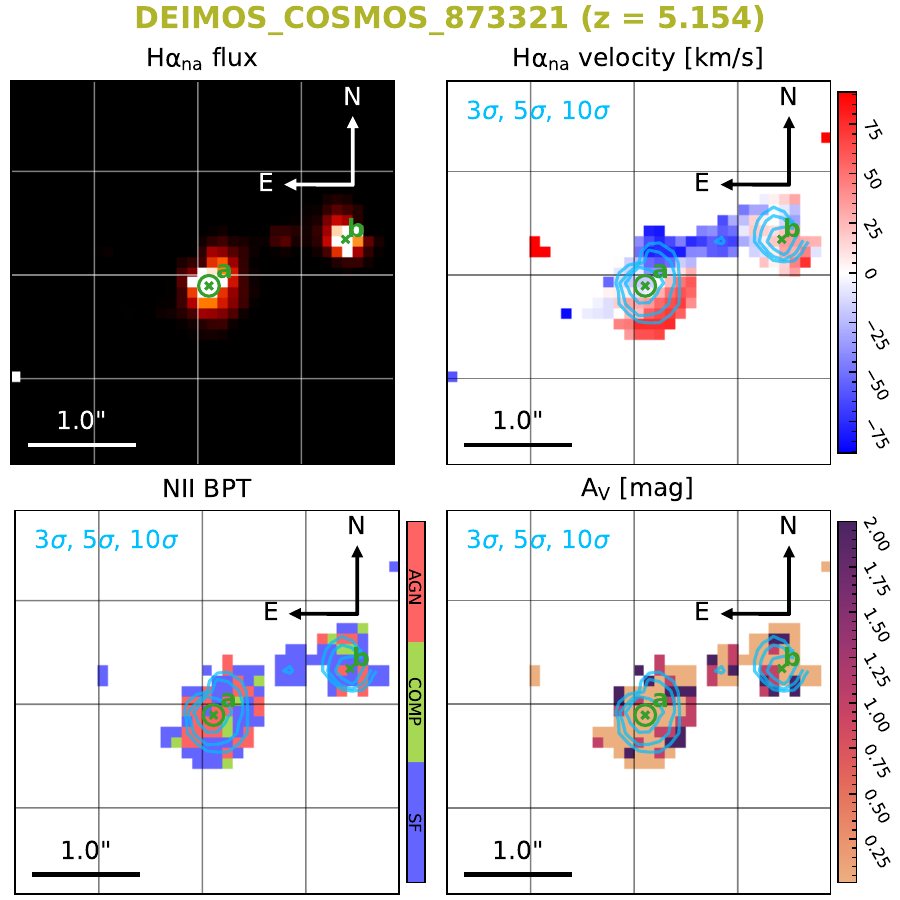}
   \end{minipage}
   \caption{Spatially resolved properties of the 7 Type-1 AGN candidates. For each target, we present: \Ha\ line flux map (top left), \Ha\ velocity map (top right), \NII-BPT diagnostic map (bottom right), and dust attenuation map (bottom left). The attenuation is derived from the Balmer decrement using the narrow component \Ha/\Hb\ ratio. Blue contours overlaid on each map show the \Ha\ line flux from 3$\sigma$ (outermost) to 20$\sigma$ (innermost, where applicable). Lime crosses mark all aperture centres analysed in this work, with circles indicating those containing AGN candidates.}
   \label{fig:AGN_morph}
\end{figure*}

From our spectral analysis, we identified 7 Type-1 AGN candidates in our ALPINE-CRISTAL-JWST sample (with two residing in a single merger system). To investigate the relationship between AGN activity and galaxy properties, we performed emission-line fitting on the IFU data cube to obtain spatially resolved maps of line flux, velocity offset and line ratios. We only apply a narrow component fitting procedure due to the limited S/N of the cube data. The results are presented in Fig. \ref{fig:AGN_morph}.

The spatial intensity distributions are generally consistent across different emission lines, thus we only present the \Ha\ line flux map in Fig. \ref{fig:AGN_morph}. The \Ha\ line flux maps reveal that four of our candidates exhibit multiple centres, likely indicating merger systems. The velocity gradients of their host galaxies are well captured in our velocity maps. We note that the velocity dispersion maps cannot be well constrained and are not displayed due to the low spectral resolution and low S/N of the cube data.

Compared to previous \CII\ kinematic analyses \citep{Jones2021,Romano2021,Posses2024}, our IFU data provide higher spatial resolution, offering more detailed insights into galaxy morphology and kinematics. Among our AGN candidates, we identify four systems with multiple centres in their \Ha\ flux maps, including three (\textit{DC\_417567}, \textit{DC\_683613}, and \textit{DC\_873321}) that were previously identified as mergers in ALPINE \citep{Jones2021,Romano2021} and CRISTAL \citep{HerreraCamus2025a} studies. The merger nature of \textit{DC\_683613} has been further confirmed by recent CRISTAL data \citep[][see Fig.3]{Posses2024}, consistent with our IFU observations.

\textit{DC\_848185} presents a particularly complex and intriguing case. The flux map reveals a clear morphological segregation between a smaller galaxy to the northwest (labelled \texttt{a}) and a larger system containing three distinct peaks (labelled \texttt{b}, \texttt{c}, and \texttt{d}) in the southeast. The weak connection between these components suggests an early-stage merger system. We identify an AGN candidate at the geometric centre of the smaller galaxy (peak \texttt{a}), where the velocity map shows only marginal rotation.

The major component exhibits more complex features. \citet[][in prep]{Davies2025} report evidence for a biconical outflow centred at aperture \texttt{d}, extending in the east-west direction. While we detect broad \Ha\ emission at position \texttt{d}, this feature fails our line flux ratio criterion (eq. \ref{eq:line_ratio_criterion}) and is likely associated with the outflow. At position \texttt{b}, however, the absence of \OIII\ outflow signatures combined with the broad \Ha\ component suggests an AGN origin. The location of \texttt{b} near the dynamical centre of the galaxy further supports this interpretation.

The southernmost region, around aperture \texttt{c}, exhibits exceptionally strong \OIIIdoublet\ emission that dominates over all other apertures, shifting these pixels towards the AGN region of diagnostic diagrams. Although we do not detect a broad component in the aperture spectrum at \texttt{c}, the line ratios suggest the presence of a Type-2 AGN, as indicated by both \NII- and \SII-BPT diagnostic diagrams.

In contrast to previous merger classifications based on \CII\ maps \citep{LeFevre2020,Romano2021}, our IFU data reveal that \textit{DC\_519281} and \textit{DC\_536534} appear as compact, isolated galaxies with well-defined velocity gradients. However, we cannot exclude the possibility that these systems may harbor additional structures obscured by heavy dust, which is revealed through long-wavelength ALMA observations \citep[e.g.,][]{Liu2024a}. In addition, while \textit{DC\_536534} shows no additional peaks in our \Ha\ map, contrary to its \CII\ morphology reported by \citet{Romano2021}, our latest stacked ALMA data reveal intriguing gas kinematics features (see Section \ref{subsec:feedback} for details).

Notably, among the 7 merger galaxies identified in the ALPINE-CRISTAL-JWST sample using \CII\ maps \citep{Romano2021}, 5 of them are identified as AGN candidates in our study. The two missed galaxies, \textit{VC\_5100541407} and \textit{VC\_5100822662}, were excluded due to insufficient overall S/N ratio in the former case and low flux in the latter, where a potential broad \Ha\ component is unconvencing due to the strong outflow signature detected in its \OIII\ lines.

In summary, our rest-frame optical morphological analysis demonstrates that five out of seven AGN candidates are associated with merger or clumpy systems, including two of them residing in a same complex merger system \textit{DC\_848185}. The remaining two candidates reside in galaxies that appear isolated with clear velocity gradients in our IFU data but are identified as mergers by ALMA observations \citep{Romano2021}. This strong correlation between AGN identification and merger/clumpy systems suggests that galaxy interactions may facilitate gas inflow and AGN fueling. However, we caution that the complex gas kinematics in these systems could potentially lead to misidentification of AGN broad components.

\subsection{Emission-line ratio diagnostics} \label{subsec:BPT}

\begin{figure*}
   \includegraphics[width=\textwidth]{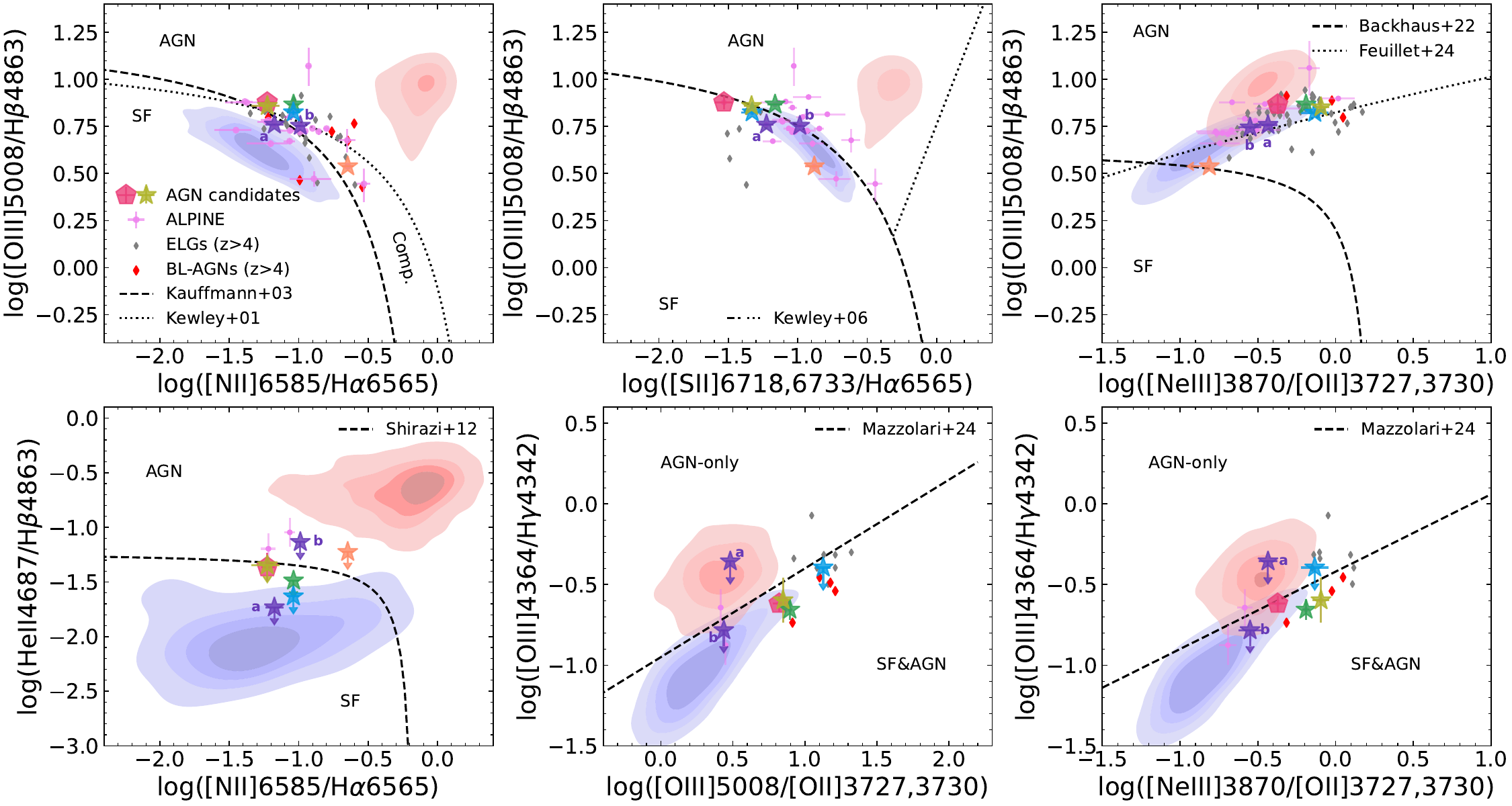}
   \caption{Emission-line ratio diagnostic diagrams for our AGN candidates and comparison samples. Top row: Classic BPT diagrams showing \OIII/\Hb\ versus \NII/\Ha\ (left) and \SII/\Ha\ (middle), and the metallicity-independent OHNO diagram showing \OIII/\Hb\ versus \NeIII/\OII\ (right) with demarcation lines from \citet{Backhaus2022} and \citet{Feuillet2024}. Bottom row: Alternative diagnostic diagrams using the high-ionization \HeII\ line \citep[left;][]{Shirazi2012} and auroral line ratios involving \OIIIARR\ and \Hg\ \citep[middle and right;][]{Mazzolari2024}. Our AGN candidates are shown as coloured stars (pentagon for the robust case), while unidentified optical peaks from our sample are shown as pink dots. Blue and red contours show the distribution of low-redshift SDSS galaxies classified as star-forming and AGN respectively using the N2-R3 BPT criteria. Gray diamonds represent high-redshift emission-line galaxies from other JWST surveys \citep[ERO, GLASS, CEERS, and JADES, ][]{Nakajima2023,DEugenio2024}, with confirmed broad-line AGNs from \citet{Harikane2023} and \citet{Maiolino2024} highlighted in red.}
   \label{fig:BPT}
\end{figure*}

To aid in narrow line diagnostics, we measured additional emission line fluxes including \HeII, \OIIIARR, \Hg, \NeIII, and \OIIdoublet, fitting each with a single Gaussian component. These measurements were performed on the same aperture spectra used for our broad-line analysis, ensuring consistent spatial sampling across all diagnostic methods. For lines with significance below 3$\sigma$, we report upper limits at the 3$\sigma$ level. Table \ref{tab:line_measurements} summarises the measured emission line fluxes used in this section.

The narrow emission lines measured from our aperture spectra enable us to place our AGN candidates on various diagnostic diagrams, e.g., the classic BPT diagram and VO87 diagrams \citep{Baldwin1981,Veilleux1987}. These two-dimensional diagnostic plots have long served as effective tools for distinguishing between star formation-powered and AGN-powered emission-line galaxies. However, at high redshift, several factors complicate their interpretation: star-forming galaxies exhibit higher \OIII/\Hb\ ratios due to increased ionization parameters, whilst AGN narrow-line regions show weaker \NII\ emission due to lower metallicity. These effects shift sources towards the canonical demarcation lines defined by \citet{Kewley2001,Kauffmann2003,Kewley2006}. In response to these challenges, alternative diagnostic methods have been proposed, including the metallicity-independent \NeIII/\OII\ ratio \citep[``OHNO'' diagram,][]{Backhaus2022,Feuillet2024}, high-ionization lines \citep[e.g., \HeII,][]{Shirazi2012, Uebler2023a, Scholtz2025}, and the strong auroral line \OIIIARR\ as a replacement for weak \NII\ and \SII\ emissions \citep{Mazzolari2024,Uebler2024}. Additionally, rest-frame UV emission line diagnostics using lines such as \CIIIlambda, \OIIIUVlambda, \CIVlambda, and \HeIIlambda\ have been developed for high-redshift galaxies \citep{Hirschmann2019}, though these lines are not covered by our current dataset.

Fig. \ref{fig:BPT} presents our AGN candidates and non-BLAGN optical centres on these diagnostic diagrams, alongside comparison samples from SDSS DR7 at low redshift \citep{Abazajian2009, Shirazi2012} and high-redshift samples from JWST ERO, GLASS, CEERS and JADES programmes \citep{Nakajima2023,DEugenio2024}. Among the comparison sample, high-redshift galaxies confirmed to host broad-line AGN \citep{Harikane2023,Maiolino2024} are highlighted in red. For our AGN candidates with emission lines below S/N$<$3, we plot 1$\sigma$ upper limits. For comparison samples, we require S/N$>$5 for low-redshift and S/N$>$3 for high-redshift sources across all relevant emission lines. To ensure robust classifications, we only plot comparison sources for which every emission line required for a given diagnostic diagram is detected. This conservative approach, which excludes sources with upper limits on any relevant lines, naturally resulting in fewer data points. The SDSS sample is classified using the \NII/\Ha\ versus \OIII/\Hb\ (N2-R3) diagram, with star-forming galaxies shown in blue contours and AGNs in red contours.

Our observational strategy ensures sufficient S/N for all AGN candidates in both the N2-R3 and S2-R3 diagrams. While most candidates are classified as AGN or composite objects, these classifications must be interpreted with caution. At high redshifts ($z \gtrsim 2$), the different physical conditions in the interstellar medium cause star-forming galaxies to occupy the ``composite'' region of the BPT diagram \citep{strom2017}. Furthermore, as noted in previous studies \citep[e.g.,][]{Harikane2023,Nakajima2023,Uebler2023a,Maiolino2024}, the low metallicity of high-redshift galaxies results in exceptionally low N2 and S2 ratios. This effect, combined with the enhancement in the degree of nebular excitation high R3 values, can cause AGN to overlap with the star-forming galaxy population \citep{Greene2007,Maiolino2019}. Consequently, the BPT diagrams alone, while useful for comparison with other studies, cannot definitively establish the AGN nature of our candidates.

Alternative diagnostic methods prove similarly ambiguous. In the metallicity-independent ``OHNO'' diagram (upper-right panel of Fig. \ref{fig:BPT}), our AGN candidates show no clear separation from other high-redshift galaxies, though they do fall within the overlap region between local star-forming galaxies and AGN.

Of our 7 AGN candidates, only \textit{DC\_536534} shows significant \HeII\ emission, lying just below the demarcation line of \citet{Shirazi2012}. The remaining candidates, having only \HeII\ upper limits, cluster near the local star-forming galaxy region. Beyond \textit{DC\_536534}, we detect \HeII\ in two other aperture centres (shown as pink dots), though all these detections barely exceed 3$\sigma$ significance. Notably, \HeII\ remains undetected in most high-redshift faint BLAGNs observed by JWST \citep{Harikane2023,Maiolino2024}, suggesting limited utility for weak AGN selection.

The positions of our candidates in the \OIIIARR\ diagnostic diagram also fall within the overlap region, clustering with other high-redshift BLAGNs. \textit{DC\_683613} is absent from the last two panels due to insufficient S/N of both \Hg\ and \OIIIARR\ lines. Despite the clear separation between local star-forming galaxies and AGNs, the \OIIIARR\ diagnostic proves inconclusive for our candidates.

In summary, none of these diagnostic methods definitively identifies our candidates as AGN. We interpret this ambiguity as intrinsic to faint AGN, which likely occupy a genuine composite region where both AGN and star-formation processes contribute significantly to the emission lines. These weak AGN consistently populate the intersection between star-forming and AGN populations across all diagnostic diagrams.

\section{Discussion} \label{sec:discussion}

\subsection{Fraction of AGN} \label{subsec:AGN_Fraction}

Our analysis has identified 7 Type-1 AGN candidates among the 33 photometric centres examined across 18 ALPINE-CRISTAL-JWST galaxies. Of these candidates, one presents as a robust detection, exhibiting both typical broad-line width and clear spatial distribution. Two candidates are found within merger systems. We exclude \textit{DC\_873756} from our final sample due to its ambiguous AGN nature and low spectral S/N (see Appendix \ref{Appendix}). We also omit \textit{DC\_842313}, which contains three optical centres, due to the lack of G395M coverage necessary for \Ha\ analysis.

These results yield an AGN fraction ranging from 5.9\% (considering only the robust detection) to 33\% (including all candidates). This range can be contextualized within the broader landscape of high-redshift AGN studies. Previous BLAGN surveys report detection rates of 1-11\% \citep{Harikane2023,Maiolino2024,Matthee2024}, while studies of Type-2 AGNs at comparable redshift find rates around 20\% \citep{Scholtz2025,Treiber2025}. Our results align more closely with the 10-35\% fraction reported by \citet{Fujimoto2024}. Notably, all observationally determined AGN fractions remain substantially below the expected 50\% from cosmological simulations \citep{Baker2025}.

While the complexity of the ALPINE-CRISTAL-JWST sample selection precludes definitive conclusions about AGN fraction evolution, several factors may explain our possible elevated detection rate.

The stellar mass threshold of our sample likely plays a crucial role. As discussed in Sec. \ref{sec:detectability}, broad emission lines from AGN in low-mass galaxies are inherently difficult to detect, even allowing for potential evolutionary effects. AGN with $M_{\rm BH}<10^6~M_\odot$ typically produce broad lines with widths near or below 1000 \kms, making them challenging to distinguish from outflows or instrumental broadening. Our $M_\star$ selection ensures more readily detectable broad-line signatures, enhancing our identification efficiency.

Our IFU data offer several advantages over spatially unresolved spectroscopic approaches. The spatially resolved data enable us to focus specifically on galaxy centres, minimizing contamination from large-scale outflows and reducing noise levels. In addition to the subtraction of recognizable \NII\ and \SII\ doublets, we can better deblend the broad \Ha\ component from the host galaxy emission.

The high incidence of merger systems in our sample likely contributes to our enhanced AGN detection rate. As detailed in Section \ref{subsec:morphology}, all identified AGN candidates are associated with merger or clumpy structures, as evidenced by either our IFU \Ha\ maps or previous ALMA observations. While debate continues regarding whether mergers enhance AGN activity \citep[e.g.,][]{Silva2021,Secrest2020,Micic2024}, each merger centre likely hosts a supermassive black hole. Our comprehensive examination of all potential nuclei within these systems has significantly increased our detection probability. Notably, only two candidates are found in isolated galaxies, and the identification of dual AGN candidates in \textit{DC\_848185} demonstrates the advantage of our approach over conventional single-slit spectroscopy.

Finally, we note an important consideration regarding the redshift distribution of our AGN candidates. While 10 of our 18 target galaxies lie at $z<5$, all confirmed AGN candidates are located at $z>5$. This apparent concentration at higher redshifts likely represents a selection bias rather than an intrinsic evolutionary trend. A few potential AGNs at $z<5$ were excluded from our final sample due to technical limitations. For instance, the galaxy \textit{DC\_873756} ($z=4.54$) likely hosts a broad-line AGN, but its low S/N ratio prevents definitive confirmation (see Appendix \ref{Appendix}). Similarly, \textit{DC\_842313} ($z=4.55$) shows evidence of a hidden AGN \citep{Solimano2025}, but lacks the necessary G395M coverage for \Ha\ analysis. Furthermore, the over-representation of Ly$\alpha$ emitters at $z>5$ in our DEIMOS sample \citep{LeFevre2020} may contribute to the observed redshift distribution of our AGN candidates.

\subsection{Gas Kinematics of \textit{DC\_536534}} \label{subsec:feedback}

\begin{figure}
   \includegraphics[width=\columnwidth]{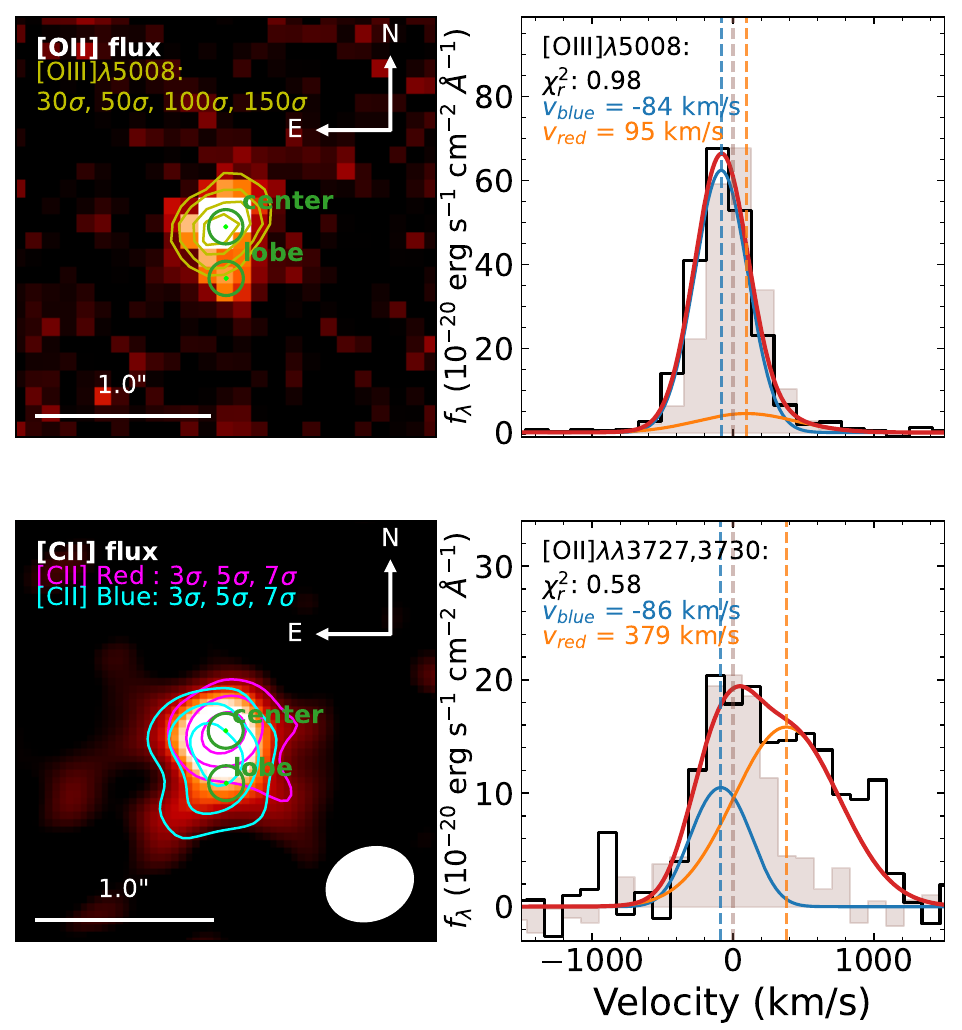}
	\caption{Extended emission features in our most robust AGN candidate \textit{DC\_536534}. Left panels: Emission-line maps showing the southern lobe structure. The upper panel presents the \OII\ line flux map with \OIII\,$\lambda$5007 flux contours overlaid in yellow. The lower panel shows the \CII\ emission map, revealing an extended lobe south-east of the galaxy centre. Magenta and cyan contours represent the red- and blue-shifted \CII\ emission (integrated $\pm$430 \kms\ from the line centre), suggesting outflow kinematics. Contour levels are drawn at 3$\sigma$, 5$\sigma$, and 7$\sigma$ (1$\sigma=0.018$ Jy beam$^{-1}$ \kms). Right panels: \OIII\ and \OII\ spectra extracted from the apertures marked in the left panels. Central emission (shaded) is normalized to the peak of the lobe spectrum (solid steps). The spectra of the lobe are fitted with two Gaussian components. The reduced $\chi^2$ of the fit and the velocity offsets of the two components, denoted as $v_{\text{red}}$ and $v_{\text{blue}}$, are indicated in the top-left corner of the spectral plots.}
	\label{fig:AGN_Dynamic}
\end{figure}

Our most convincing AGN candidate, \textit{DC\_536534}, exhibits complex gas kinematics beyond simple rotation. While the \Ha\ and \OIII\ velocity maps (Fig. \ref{fig:AGN_morph}) reveal clear rotational patterns in an apparently isolated galaxy, the \OII\ emission line map (upper-left panel of Fig. \ref{fig:AGN_Dynamic}) shows an additional lobe structure extending southward from the galaxy centre. This lobe structure is detected at $>5\sigma$ significance, ruling out the effect of noise.

To investigate the kinematics of this lobe, we extracted spectra from two apertures: one centred on the galaxy core and another offset 0.3\arcsec\ southward, each with a radius of 0.1\arcsec. The right panels of Fig. \ref{fig:AGN_Dynamic} compare the \OIII\ and \OII\ line profiles from these apertures, with the central emission (shaded) normalised to match the peak of the lobe spectrum (solid steps). The lobe spectrum exhibits a pronounced red wing in \OII, corresponding to the observed lobe structure. In contrast, the \OIII\ line profile in the lobe appears largely consistent with the central emission, showing only a slight blueshift attributable to galaxy rotation.

Our spectral analysis of the lobe employs a two-component Gaussian model. Both \OIII\ and \OII\ lines show a blueshifted narrow component with $v_{\rm off}\simeq85~\kms$, consistent with the rotation pattern of the galaxy. The \OII\ line additionally exhibits a redshifted component with $v_{\rm off}=379~\kms$, approximately three times brighter than the core component. However, this complex profile of the redshifted feature prevents accurate modelling with a single Gaussian. A similar but much fainter redshifted component appears in the \OIII\ line, contributing only 13\% of the flux of the core component. The detection of these features in both gratings (\OII\ in the G235M and \OIII\ in the G395M) confirms their physical origin rather than instrumental effects.

The line ratios of the two fitted components differ markedly, with $\log{f_{[OII]}/f_{[OIII]}}_{\rm blue} = -0.86$ for the blue core and $\log{f_{[OII]}/f_{[OIII]}}_{\rm red} = 0.44$ for the redshifted component. Despite this substantial difference, both ratios are consistent with the broad range observed in other samples \citep[e.g.,][]{Silverman2009}, which allows for the interpretation that the lobe is a genuine astrophysical feature rather than an artifact of our spectral decomposition.

To complement our analysis of ionized gas, we examined the neutral gas distribution through \CII\ observations of \textit{DC\_536534}. We combined data from ALPINE \citep{Bethermin2020} and CRISTAL \citep{HerreraCamus2025a} programs to achieve enhanced dynamic range. Following continuum subtraction, we generated spectral cubes using \textit{tclean} in two modes: \textit{cube} mode with 20 km/s channels and \textit{mfs} mode integrating over 2$\times$ FWHM from \citet{Bethermin2020}. The final \CII\ map, produced with robust Briggs weighting yielding a 0.47\arcsec\ beam \citep{1995AAS...18711202B}, is presented in the lower-left panel of Fig. \ref{fig:AGN_Dynamic} \citep[see ][for detailed reduction procedures]{Liu2025}. The map reveals an additional lobe extending south-eastward from the galaxy centre. We generated velocity-divided contours by integrating $\pm430~\kms$ (3$\sigma$) from the line centre to the red- and blue-shifted components, shown in magenta (redshifted) and cyan (blueshifted). While these contours align with the rotation axis derived from ionized gas, the kinematics of the lobe are enhanced by blueshifted signatures.

The \CII\ emission of \textit{DC\_536534}, which traces extensive neutral gas structures with minimal obscuration, reveals a distinct lobe extending southeast from the galaxy center. The kinematics of this feature are broadly aligned with the rotation, providing evidence that the structure is a tidal feature from an ongoing minor merger \citep{Villalobos2008}. At the base of this structure, approximately 1.5 kpc south of the center, we detect a redshifted component in both \OII\ and \OIII\ lines. This component is characterized by strong \OII\ flux but a significantly weaker \OIII\ signal. The discrepancy between the bulk gas motion (traced by \CII) and the ionized gas (traced by \OII) can be naturally explained by a moderate-velocity shock front, which heated the interstellar medium and result in the intense, low-ionization \OII\ emission and its unique velocity signature \citep{Rich2011}. An alternative, or perhaps complementary, scenario is that the interaction has triggered a localized burst of star formation in this region, as evident by the low-ionization \OII\ emission.

\subsection{Stellar Mass Estimation} \label{subsec:Mstar_Estimation}

\begin{figure}
   \includegraphics[width=\columnwidth]{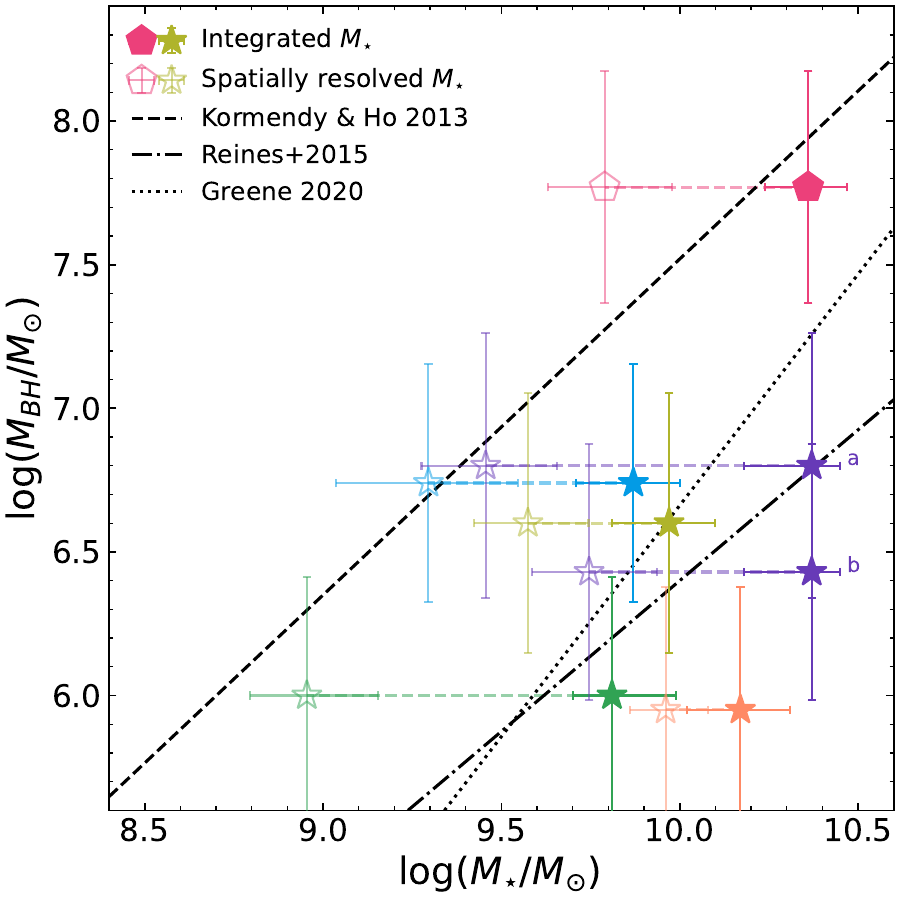}
   \caption{Comparison of black hole-stellar mass relations using different stellar mass estimation methods. Filled symbols show measurements using integrated broad-band photometry from \citet{Faisst2020}, which are adopted as fiducial values for this study. Open symbols represent spatially resolved estimates from JWST data \citep[][submitted]{Tsujita2025}. Dashed lines connect measurements of the same object. Other lines are identical to Fig. \ref{fig:MM_relation}.}
	\label{fig:MM_relation_pix}
\end{figure}

The stellar mass estimation used in this work is derived through SED fitting with pre-JWST broad-band photometry from \citet{Faisst2020}. These spatially integrated measurements can be affected by strong emission lines in actively star-forming galaxies, potentially bias the stellar mass estimation. To explore this issue, we analysed the $M_{\rm BH}-M_\star$ relations using spatially resolved stellar mass measurements from \citet[][submitted]{Tsujita2025}. Their analysis employs {\tt Prospector} to simultaneously fit NIRCam photometry and NIRSpec IFU spectroscopy at $\sim$0.6 kpc resolution. For each AGN candidate, we summed the stellar masses of constituent pixels, considering only pixels within the host galaxy for merger systems and corrected the flux loss due to the mask of low S/N pixels. 

As shown in Fig. \ref{fig:MM_relation_pix}, for our AGN candidate sample, these spatially resolved estimates are systematically lower by $\sim$0.6 dex compared to the integrated broad-band measurements. It is worth noting that the spatially resolved stellar masses for the full ALPINE sample are also presented in \citet[][submitted]{Tsujita2025}. The general deviation for the 18 ALPINE galaxies is only $\sim$0.35 dex on average. This indicates a more significant discrepancy in our AGN candidates compared to the general galaxy population at similar redshifts.

The existence of AGNs and their features in the SED could be a key factor introducing this bias \citep{Cardoso2017}. This effect could be particularly important at shorter wavelengths where the featureless power-law continuum of AGNs can dominate. To quantify this contamination, we estimated the AGN contribution at rest-frame 1350\AA\ using the attenuation-corrected broad H$\alpha$ luminosity following \citet{Jun2015}. Assuming a typical AGN spectral slope ($f_{\lambda}\propto\lambda^{-2.08}$), we predict GALEX FUV magnitudes 2-4 magnitudes fainter than the total emission measured by \citet{Faisst2020}, suggesting minimal AGN contamination in our stellar mass estimates.

Additional uncertainty could arise from strong narrow-line emission. Another spatially resolved SED fitting result based solely on the broad-band images \citep{Li2024e} finds no offsets between the $M_{\rm BH}$ derived from the integrated measurements regardless the different SED models. Further discussion about the origin of these differences are beyond the scope of this study, which will be further addressed in \citet[][submitted]{Tsujita2025}.

While spatially-resolved measurements yield lower stellar masses and shift the position of our AGN candidates on the $M_{\rm BH}-M_\star$ plane, our conclusions remain unchanged. These new AGN candidates are undermassive, when comparing to the \citet{Kormendy2013} relation, and scattered around the relations of \citet{Greene2020} and \citet{Reines2015}.

\section{Summary} \label{sec:summary}

We conduct a systematic search for broad-line AGNs in the ALPINE-CRISTAL-JWST sample, comprising 18 star-forming main-sequence galaxies with $M_\star>10^{9.5}M_\odot$ at $z=4.4-5.7$. Using NIRSpec IFU data, we examine 33 photometric centres and identify 7 Type-1 AGN candidates. Our key findings are as follows:

\begin{enumerate}
   \item We identify one highly convincing AGN candidate, \textit{DC\_536534}, exhibiting a broad \Ha\ component with FWHM $\sim$2800 \kms. This broad line is well separated from both the narrow core and an outflow component, detected in \OIII\ lines and exhibits a PSF-like spatial profile. \textit{DC\_536534} shows a significant high-ionization \HeII\ line (S/N $>$ 3), not seen in other targets.
   
   \item Including six additional candidates with broad \Ha\ components (FWHM $\simeq 600-1600$ \kms), we find an AGN fraction ranging from 5.9\% (considering only the most robust case) to 33\%. We attribute this relatively high fraction to our focus on high-mass galaxies, where broad-line detection is more feasible, combined with the advantages of spatially resolved IFU data and our optimised observational strategy.
   
   \item Through detailed simulations, we demonstrate the inherent challenges in detecting faint AGN in low-mass galaxies. For galaxies with $M_\star<10^{9.5}M_\odot$, the expected broad-line widths and luminosities of the local-relation predicted AGNs fall below current detection thresholds. This selection effect helps explain the apparent prevalence of overmassive black holes relative to host galaxies in previous high-redshift studies.
  
   \item Unlike previous JWST AGN studies at high redshift that found overmassive black holes relative to their host galaxies, our candidates lie either close to or below the local $M_{\rm BH}-M_\star$ scaling relations. This demonstrates that when accounting for mass-dependent selection biases, AGNs following local scaling relations can be detected at high redshift.
   
   \item Our morphological analysis reveals a strong association between AGN activity and galaxy interactions, with five of our seven AGN candidates residing in merger or clumpy systems, including two candidates found within the same complex merger system (\textit{DC\_848185}). Even the two candidates that appear as isolated galaxies in our IFU data have been identified as mergers in previous ALMA observations. This preference suggests that galaxy interactions may facilitate gas inflows that trigger AGN activity at high redshift.
   
   \item The emission-line ratios of our AGN candidates consistently lie within narrow emission-line ratio diagnostic diagrams where star-forming galaxies and AGNs overlap. This pattern persists across multiple diagnostic methods, including classic BPT diagrams, the metallicity-independent ``OHNO'' diagram, and diagnostics employing high-ionization or auroral lines. This suggests these weak AGNs may possess an intrinsically composite nature, with significant contributions from both AGN and star-formation processes.
\end{enumerate}

In conclusion, our study provides new insights into AGN-host galaxy co-evolution at high redshift by identifying faint broad AGN components in a well-defined galaxy sample. We highlight the critical importance of considering mass-dependent selection biases when studying AGN scaling relations. Current emission-line ratio diagnostics prove insufficient for definitive high-redshift AGN identification. Future progress may come through enhanced detection of high-ionization lines via improved spectral sensitivity.

\begin{landscape}
\begin{table}
   \centering
      \begin{threeparttable}
   \caption{Properties measurements for the AGN candidates and their host galaxies in the ALPINE-CRISTAL-JWST sample\label{tab:AGN_prop}}
   \begin{tabular}{lccccccccccc}
   \toprule
   Name & RA & DEC & $z^a$ & \Ha$_{\rm br, corr}^{b}$ & ${\rm FWHM}_{\rm H\alpha,br,corr}^{c}$ & $\log(M_*)^{a}$ & $\log(M_\mathrm{BH})^{d}$ & $\log(L_\mathrm{bol})$ & $\log(L/L_\mathrm{Edd})$ & E(B-V) & $A_V$ \\
   -- & deg & deg & -- & $10^{-18}$ \ergscm & km s$^{-1}$ & $M_\odot$ & $M_\odot$ & erg s$^{-1}$ & -- & mag & mag \\
   \midrule
   DC\_417567 & 150.5170078 & 1.9289283 & 5.6700 & $1.69 \pm 0.42$ & $596 \pm 60$ & $9.81_{-0.11}^{+0.18}$ & $6.00 \pm 0.10$ & $44.15 \pm 0.09$ & $0.05$ & 0.00 & 0.00 \\
   DC\_519281 & 149.7537227 & 2.0910165 & 5.5759 & $4.40 \pm 0.35$ & $1110 \pm 132$ & $9.87_{-0.16}^{+0.13}$ & $6.74 \pm 0.11$ & $44.49 \pm 0.03$ & $-0.35$ & 0.08 & 0.32 \\
   DC\_536534 & 149.9718537 & 2.1181852 & 5.6886 & $11.69 \pm 1.43$ & $2812 \pm 157$ & $10.36_{-0.12}^{+0.11}$ & $7.78 \pm 0.06$ & $44.88 \pm 0.05$ & $-1.01$ & 0.24 & 0.98 \\
   DC\_683613 & 150.0392432 & 2.3372033 & 5.5420 & $0.80 \pm 0.14$ & $678 \pm 112$ & $10.17_{-0.15}^{+0.14}$ & $5.95 \pm 0.15$ & $43.84 \pm 0.07$ & $-0.21$ & 0.00 & 0.00 \\
   DC\_848185\_a & 150.0895382 & 2.5865496 & 5.2931 & $1.25 \pm 0.20$ & $1618 \pm 417$ & $10.37_{-0.19}^{+0.08}$ & $6.80 \pm 0.23$ & $43.97 \pm 0.06$ & $-0.93$ & 0.07 & 0.29 \\
   DC\_848185\_b & 150.0896631 & 2.5863339 & 5.2931 & $2.11 \pm 0.33$ & $955 \pm 206$ & $10.37_{-0.19}^{+0.08}$ & $6.43 \pm 0.20$ & $44.17 \pm 0.06$ & $-0.36$ & 0.20 & 0.81 \\
   DC\_873321 & 150.0169147 & 2.6266375 & 5.1542 & $4.27 \pm 0.61$ & $986 \pm 231$ & $9.97_{-0.16}^{+0.13}$ & $6.59 \pm 0.21$ & $44.41 \pm 0.05$ & $-0.28$ & 0.15 & 0.62 \\
   \bottomrule
   \end{tabular}
   \begin{tablenotes}
      \item \textbf{Note. }
      \item $^a$ The redshift is measured from far-infrared [CII] line and the $M_\star$ is derived from SED fitting (see Sec. \ref{subsec:MBH_Mstar}). Both values are taken from \citet{Faisst2020}.
      \item $^b$ The broad H$\alpha$ line flux is corrected for the dust extinction using the Balmer decrement.
      \item $^c$ The FWHM of the broad H$\alpha$ line is corrected for the instrumental broadening.
      \item $^d$ The errors for $M_{\rm BH}$ is only the statistical uncertainty propagated from the uncertainties for broad H$\alpha$ line flux and FWHM measurements.
   \end{tablenotes}
   \end{threeparttable}
\end{table}

\begin{table}
	\centering
	\begin{threeparttable}
	\caption{Emission line flux measurements for the AGN candidates in the ALPINE-CRISTAL-JWST sample\label{tab:line_measurements}}
	\begin{tabular}{lccccccccccc}
	\toprule
   Name & \SIIdoublet\ & \NII\,$\lambda$6585 & \Ha$_{\rm na}$ & \OIII\,$\lambda$5008 & \Hb$_{\rm na}$ & \HeII & \OIII\,$\lambda$4364 & \Hg\ & \NeIII\ & \OIIdoublet\ \\
	\multicolumn{11}{l}{Units ($10^{-18}$ \ergscm)} \\
	\midrule
   DC\_417567 & $0.42 \pm 0.05$ & $0.56 \pm 0.04$ & $6.09 \pm 0.42$ & $19.20 \pm 0.10$ & $2.61 \pm 0.01$ & $<0.08$ & $0.24 \pm 0.03$ & $1.07 \pm 0.05$ & $1.58 \pm 0.06$ & $2.44 \pm 0.10$ \\
DC\_519281 & $0.47 \pm 0.08$ & $0.93 \pm 0.08$ & $10.12 \pm 0.33$ & $26.24 \pm 0.87$ & $3.33 \pm 0.10$ & $<0.09$ & $<0.29$ & $0.71 \pm 0.19$ & $1.36 \pm 0.24$ & $1.83 \pm 0.21$ \\
   DC\_536534 & $0.47 \pm 0.06$ & $0.92 \pm 0.15$ & $9.45 \pm 0.88$ & $33.05 \pm 2.24$ & $2.57 \pm 0.21$ & $0.18 \pm 0.05$ & $0.43 \pm 0.05$ & $1.76 \pm 0.05$ & $1.66 \pm 0.18$ & $3.79 \pm 0.20$ \\
   DC\_683613 & $0.29 \pm 0.04$ & $0.49 \pm 0.03$ & $2.20 \pm 0.15$ & $2.87 \pm 0.04$ & $0.83 \pm 0.01$ & $<0.05$ & $<0.53$ & $<0.30$ & $<0.24$ & $1.55 \pm 0.14$ \\
   DC\_848185\_a & $0.34 \pm 0.04$ & $0.38 \pm 0.05$ & $5.65 \pm 0.14$ & $10.80 \pm 0.12$ & $1.88 \pm 0.02$ & $<0.03$ & $<0.36$ & $0.82 \pm 0.14$ & $1.25 \pm 0.15$ & $3.36 \pm 0.24$ \\
   DC\_848185\_b & $0.77 \pm 0.07$ & $0.77 \pm 0.06$ & $7.49 \pm 0.21$ & $12.14 \pm 0.15$ & $2.14 \pm 0.03$ & $<0.15$ & $<0.15$ & $0.90 \pm 0.12$ & $1.03 \pm 0.16$ & $3.56 \pm 0.21$ \\
   DC\_873321 & $0.48 \pm 0.06$ & $0.61 \pm 0.11$ & $10.34 \pm 0.51$ & $26.08 \pm 1.01$ & $3.12 \pm 0.10$ & $<0.16$ & $0.33 \pm 0.10$ & $1.29 \pm 0.13$ & $2.59 \pm 0.21$ & $3.15 \pm 0.30$ \\
	\bottomrule
	\end{tabular}
	\begin{tablenotes}
		\item \textbf{Note.} For lines with significance $<3\sigma$, we provide the 3$\sigma$ level as the upper limit.
	\end{tablenotes}
	\end{threeparttable}
\end{table}
\end{landscape}

\section*{Acknowledgements}

W. R. is supported by the China Scholarship Council No. 202306340001. W. R. would like to thank Hannah Übler for very useful scientiﬁc suggestions and feedback.
V. V. acknowledges support from the ANID BASAL project FB210003. 
M. A. is supported by FONDECYT grant number 1252054, and  gratefully acknowledges support from ANID Basal Project FB210003 and ANID MILENIO NCN2024\_112. 
M. S. was financially supported by Becas-ANID scholarship \#21221511, and also acknowledges support from ANID BASAL project FB210003. 
A. N. acknowledges support from the Narodowe Centrum Nauki (NCN), Poland, through the SONATA BIS grant UMO2020/38/E/ST9/00077. 
W. W. acknowledges the grant support from NASA through JWST-GO-3950. 
G. C. J. acknowledges support by the Science and Technology Facilities Council (STFC), by the ERC through Advanced Grant 695671 ``QUENCH'', and by the UKRI Frontier Research grant RISEandFALL. 
J. M. gratefully acknowledges support from ANID MILENIO NCN2024\_112.
E. I. gratefully acknowledges financial support from ANID - MILENIO - NCN2024\_112 and ANID FONDECYT Regular 1221846.
This work is based in part on observations made with the NASA/ESA/CSA \textit{James Webb} Space Telescope. The data were obtained from the Mikulski Archive for Space Telescopes at the Space Telescope Science Institute, which is operated by the Association of Universities for Research in Astronomy, Inc., under NASA contract NAS 5-03127 for JWST. These observations are associated with program JWST-GO- 3045. Support for program JWST-GO- 3045 was provided by NASA through a grant from the Space Telescope Science Institute, which is operated by the Association of Universities for Research in Astronomy, Inc., under NASA contract NAS 5-03127.
This paper makes use of the following ALMA data: ADS/JAO.ALMA\#2017.1.00428.L, ADS/JAO.ALMA\#2021.1.00280.L. ALMA is a partnership of ESO (representing its member states), NSF (USA) and NINS (Japan), together with NRC (Canada), NSTC and ASIAA (Taiwan), and KASI (Republic of Korea), in cooperation with the Republic of Chile. The Joint ALMA Observatory is operated by ESO, AUI/NRAO and NAOJ.
Some of the data presented in this paper were obtained from the Mikulski Archive for Space Telescopes (MAST) at the Space Telescope Science Institute. The specific observations analyzed can be accessed via \url{http://dx.doi.org/10.17909/cqds-qc81}. STScI is operated by the Association of Universities for Research in Astronomy, Inc., under NASA contract NAS5–26555. Support to MAST for these data is provided by the NASA Office of Space Science via grant NAG5–7584 and by other grants and contracts.

\section*{Data Availability}

The data underlying this research are available in the article.



\bibliographystyle{mnras}
\bibliography{RefLib}{} 




\appendix
\section{Spectral analysis of \textit{DC\_873756}} \label{Appendix}

\begin{figure}
   \includegraphics[width=\columnwidth]{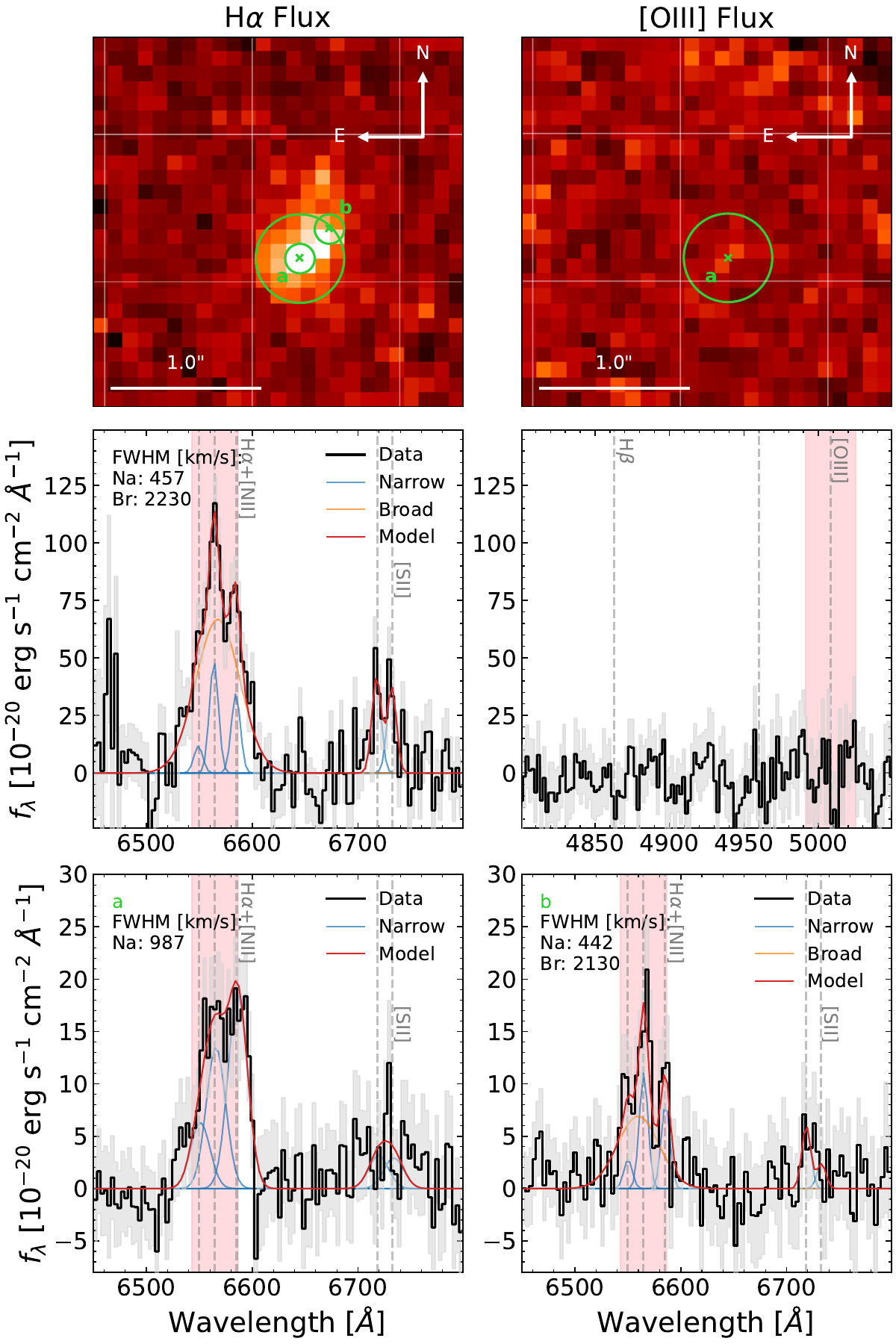}
	\caption{Emission-line flux maps and spectral analysis of \textit{DC\_873756}. Upper panels: \Ha\ and \OIII\ line flux maps, integrated over $\pm$1000 \kms\ from the line centre (integration range shown as red shading in lower panels). The lime circles indicate the 0.6\arcsec-diameter aperture used for middle panel spectral extraction and two apertures with 0.2\arcsec diameter used for the lower panels. Middle panels: spectra around \Ha\ and \Hb\ lines from the 0.6\arcsec aperture. Only the \Ha\ line is fitted with narrow$+$broad components, as neither \OIII\ nor \Hb\ is detected. Lower panels: spectra of \Ha\ line from two apertures with 0.2\arcsec diameter. We illustratively fit the \Ha\ profile from aperture \texttt{a} with ``narrow'' component and \texttt{b} with narrow$+$broad component.}
	\label{fig:Appendix_AGN}
\end{figure}

While our primary analysis focuses on confirmed AGN candidates, we present here the case of \textit{DC\_873756}, which exhibits intriguing broad emission features but lacks sufficient evidence for definitive AGN classification. The lower left panel of Fig. \ref{fig:Appendix_AGN} shows the aperture spectra from our standard analysis procedure. This spectrum seemingly reveals extremely broad \Ha\ and \NII\ lines with FWHM$\sim$1000 \kms. Since the width of these components significantly exceed the upper limit of our narrow component, our selection criteria indicate the presence of broad \Ha\ line. 

To further investigate this ambiguous case, we extracted a spectrum using a larger 0.6\arcsec-diameter aperture (compared to our standard 0.2\arcsec). This approach serves two purposes: (1) it increases the signal by capturing more flux, and (2) it potentially improves our ability to deblend narrow and broad components, particularly important for \textit{DC\_873756} where the broad component appears dominant. This larger aperture also allows us to verify the non-detection of \OIII\ and \Hb\ lines observed in our default aperture.

The upper and middle panels of Fig. \ref{fig:Appendix_AGN} present the line flux maps and larger aperture spectra. The \Ha\ map shows a clear central peak, while the \OIII\ map reveals no significant emission. When modeling the \Ha\ profile with narrow$+$broad components, we derive a black hole mass estimate of $\log{M_{\rm BH} / M_\odot}= 7.7$ with an Eddington ratio of $\log{\lambda_{\rm Edd}}=-0.74$. Combined with its stellar mass ($\log{M_\star / M_\odot} = 10.3$), these parameters place it near our robust AGN candidate \textit{DC\_536534} on the $M_{\rm BH}-M_\star$ diagram.

Interestingly, \textit{DC\_873756} exhibits relatively high metallicity indicators (N2$=-0.15$, S2$=0.22$) compared to other galaxies in our sample. These line ratios would support an AGN classification on the BPT diagram, even without O3 measurements, making this target particularly intriguing.

However, several factors prevent us from confidently classifying \textit{DC\_873756} as an AGN:

\begin{enumerate}
    \item Reliable spectral fitting is only achievable with the larger aperture; our standard aperture yields complex, blended profiles that resist meaningful decomposition.
    
    \item The broad \Ha\ component could potentially arise from complex outflows rather than an AGN, as suggested by the high velocity dispersion observed in the \CII\ line (FWHM $\sim$ 700 \kms) from ALMA data.
    
    \item The absence of typically strong \OIII\ lines is puzzling for an AGN interpretation, though this could potentially result from instrumental effects in the G235M grating band.
\end{enumerate}

To better understand the spatial distribution of emission features, we examined two standard aperture spectra (0.2\arcsec) from different regions, shown in the lower panels of Fig. \ref{fig:Appendix_AGN}. The central aperture (\texttt{a}) reveals extremely broad \Ha\ and \NII\ lines (FWHM $\sim$ 1000 \kms), potentially indicating strong outflows. However, the absence of an expected comparably broad \SII\ doublet features challenges this outflow interpretation. In the northeast region (aperture \texttt{b}), we detect a clearer set of narrow emission lines, though additional broad components are still required to fit the profile. These spatially distinct components likely contribute to the complex broad \Ha\ profile observed in the larger aperture spectrum.

Given these uncertainties and the complex nature of this source, we exclude \textit{DC\_873756} from our final AGN candidate list. Follow-up observations with higher sensitivity and spectral resolution will be necessary to definitively determine the nature of this intriguing system.

\section*{Affiliations}
$^{1}$Kavli Institute for the Physics and Mathematics of the Universe, The University of Tokyo, Kashiwa, Japan 277-8583 (Kavli IPMU, WPI)\\
$^{2}$School of Astronomy and Space Science, University of Science and Technology of China, Hefei 230026, China\\
$^{3}$Department of Astronomy, School of Science, The University of Tokyo, 7-3-1 Hongo, Bunkyo, Tokyo 113-0033, Japan\\
$^{4}$Center for Data-Driven Discovery, Kavli IPMU (WPI), UTIAS, The University of Tokyo, Kashiwa, Chiba 277-8583, Japan\\
$^{5}$Center for Astrophysical Sciences, Department of Physics \& Astronomy, Johns Hopkins University, Baltimore, MD 21218, USA\\
$^{6}$Caltech/IPAC, 1200 E. California Blvd. Pasadena, CA 91125, USA\\
$^{7}$Department of Astronomy, The University of Texas at Austin, Austin, TX, USA\\
$^{8}$Cahill Center for Astrophysics, California Institute of Technology, Pasadena, CA 91125, USA\\
$^{9}$Caltech Optical Observatories, California Institute of Technology, Pasadena, CA 91125, USA\\
$^{10}$Université Paris-Saclay, Université Paris Cité, CEA, CNRS, AIM, F-91191 Gif-sur-Yvette, France\\
$^{11}$Institute of Astronomy, Graduate School of Science, The University of Tokyo, 2-21-1 Osawa, Mitaka, Tokyo 181-0015, Japan\\
$^{12}$Instituto de Estudios Astrofísicos, Facultad de Ingeniería y Ciencias, Universidad Diego Portales, Av. Ejército 441, Santiago 8370191, Chile\\
$^{13}$Millenium Nucleus for Galaxies (MINGAL)\\
$^{14}$Centre for Astrophysics and Supercomputing, Swinburne University of Technology, Hawthorn, Victoria 3122, Australia\\
$^{15}$Sterrenkundig Observatorium, Ghent University, Krijgslaan 281 - S9, B-9000 Gent, Belgium\\
$^{16}$Département d'Astronomie, Université de Genève, Chemin Pegasi 51, 1290 Versoix, Switzerland\\
$^{17}$Departamento de Astronomía, Universidad de Concepción, Barrio Universitario, Concepción, Chile\\
$^{18}$Instituto de Física y Astronomía, Universidad de Valparaíso, Avda. Gran Bretaña 1111, Valparaíso, Chile\\
$^{19}$Kavli Institute for Cosmology, University of Cambridge, Madingley Road, Cambridge CB3 0HA, UK\\
$^{20}$Cavendish Laboratory, University of Cambridge, 19 JJ Thomson Avenue, Cambridge CB3 0HE, UK\\
$^{21}$Laboratory for Multiwavelength Astrophysics, School of Physics and Astronomy, Rochester Institute of Technology, 84 Lomb Memorial Drive, Rochester, NY 14623, USA\\
$^{22}$Space Telescope Science Institute, 3700 San Martin Drive, Baltimore, MD 21218, USA\\
$^{23}$Department for Astrophysical \& Planetary Science, University of Colorado, Boulder, CO 80309, USA\\
$^{24}$National Centre for Nuclear Research, ul. Pasteura 7, 02-093 Warsaw, Poland\\
$^{25}$INAF - Osservatorio astronomico d'Abruzzo, Via Maggini SNC, 64100, Teramo, Italy\\
$^{26}$Dept. Física Te\'{o}rica y del Cosmos, Campus de Fuentenueva, Edificio Mecenas, Universidad de Granada, E-18071, Granada, Spain\\
$^{27}$Instituto Universitario Carlos I de Física Te\'{o}rica y Computacional, Universidad de Granada, 18071, Granada, Spain\\
$^{28}$Max-Planck-Institut für Radioastronomie, Auf dem Hügel 69, 53121 Bonn, Germany\\
$^{29}$INAF - Osservatorio Astronomico di Padova, Vicolo dell'Osservatorio 5, I-35122 Padova, Italy\\
$^{30}$Institute for Astronomy, University of Hawaii, 2680 Woodlawn Drive, Honolulu, HI 96822, USA\\
$^{31}$Scuola Internazionale Superiore Studi Avanzati (SISSA), Physics Area, Via Bonomea 265, 34136 Trieste, Italy\\
$^{32}$INAF – Osservatorio di Astrofisica e Scienza dello Spazio di Bologna, Via Gobetti 93/3, 40129 Bologna, Italy

\bsp	
\label{lastpage}
\end{document}